\documentclass[10pt]{iopart}
\usepackage{iopams}
\usepackage{graphicx}    
\usepackage{rotating}
\begin{document}
\setlength{\marginparwidth}{4cm}
\def\be{\begin{equation}}
\def\ee{\end{equation}}
\def\bea{\begin{eqnarray}}
\def\eea{\end{eqnarray}}
\def\d{\mbox{d}}
\def\e{\mathop{\rm e}}
\def\p{\partial}
\def\Sds{Schwarzschild--de~Sitter}

\title[Pseudo-Newtonian and general relativistic tori in SdS spacetimes]
{Pseudo-Newtonian and general relativistic barotropic tori in Schwarzschild-de Sitter spacetimes}

\author{Zden\v{e}k Stuchl\'{i}k, Petr Slan\'{y}\footnote[1]{petr.slany@fpf.slu.cz} and Ji\v{r}\'{i} Kov\'{a}\v{r}}

\address{Institute of Physics, Faculty of Philosophy and Science,\\Silesian University in Opava, Bezru\v{c}ovo n\'{a}m. 13, Opava, CZ-746 01,\\Czech Republic}

\begin{abstract}
Pseudo-Newtonian gravitational potential introduced in spherically symmetric black-hole spacetimes with a repulsive cosmological constant is tested for equilibrium toroidal configurations of barotropic perfect fluid orbiting the black holes.
Shapes and potential depths are determined for the marginally stable barotropic tori with uniform distribution of specific angular momentum, using both the pseudo-Newtonian and fully relativistic approach. For the adiabatic (isoentropic) perfect fluid, temperature profiles, mass-density and pressure profiles and total masses of pseudo-Newtonian and relativistic tori are compared providing important information on the relevance of the test-disc approximation in both the approaches. It is shown that the pseudo-Newtonian approach can be precise enough and useful for the modelling of accretion discs in the Schwarzschild-de Sitter spacetimes with the cosmological parameter $y=\Lambda M^2/3\lesssim 10^{-6}$.
For astrophysically relevant black holes with $y<10^{-25}$, this statement is tested and shown to be precise in few percent for both accretion and excretion tori and for the marginally bound, i.e., maximally extended tori allowing simultaneous inflow to the black hole and outflow to the outer space.
\end{abstract}
\pacs{04.20.--q, 04.25.--g, 04.70.--s, 95.30.Lz, 98.62.Mw, 98.80.Es}

\section{Introduction}
\label{intro}
Cosmological observations of distant Ia-type supernova explosions indicate an~accelerating universe. Starting at the cosmological redshift $z\approx 1$, the accelerated expansion should be generated by some appropriate form of the so-called dark energy \cite{Per-etal:1999:ASTRJ2:,Rie-etal:2004:ASTRJ2:}. These results are in accord with a large variety of cosmological tests including gravitational lensing, galaxy number counts, etc. \cite{Ost-Ste:1995:NATURE:}. The recent detailed studies of the cosmic microwave background (CMB) anisotropies indicate that the energy content of the dark energy represents $\sim 74.5\%$ of the energy content in the observable universe, and the sum of energy densities is very close to the critical energy density $\rho_{\rm crit}$, corresponding to almost flat universe \cite{Spe-etal:2003:ASTJS:,Spe-etal:2007:ASTJS:}.

A large variety of possible candidates for the dark energy is discussed in these days. First of all, there is the standard possibility represented by the cosmological constant $\Lambda$. Its Lorentz invariant form enables interpretation in terms of a~ground state or vacuum energy of quantum fields \cite{Kol-Tur:1990:FrontiersinPhysics:,Dol-Zel-Saz:1988:IzdatMoskUniv:}. The energy density $\rho_{\Lambda}$, which can be associated with the cosmological constant, remains unchanged during the cosmic expansion, and its pressure to energy density ratio (equation of state) is $w=p_{\Lambda}/\rho_{\Lambda}=-1$. 
Further, there is a variety of scalar fields evolving outside of their energy minimum, called quintessence, which possess a time varying energy density and equation of state with $w<-1/3$ \cite{Zla-Wan-Ste:1999:PHYRL:}. Such a scenario can be realised by light scalar field coming from modified $f(R)$ gravity \cite{Noj-Odi:2003:PHYSR4:}, string-inspired cosmologies \cite{Tsu-Sam:2007:JOCAP:}, cosmology with extra dimensions \cite{Neu:2004:CLAQG:}, or by $k$-essence being a scalar field with a non-canonical kinetic term \cite{Arm-Dam-Muk:1999:PHYLB:,Gar-Muk:1999:PHYLB:}. Similar behaviour is exhibited by the coupled dark energy, i.e., a scalar field coupled to the dark matter. For example, Chaplygin gas or its generalization called quartessence explain both dark energy and dark matter from an unified physical origin \cite{Kam-Mos-Pas:2001:PHYLB:}. The holographic dark energy treats vacuum energy limited by a holographic conception on degrees of freedom \cite{Li:2004:PHYLB:}. Modified-gravity origin of the cosmic acceleration comes also from multidimensional theories of the braneworld-type cosmology \cite{Ger:2008:PHYSR4:}. All of these models have the equation of state with the parameter $w$ varying during the cosmological expansion. 
Another possibility is the evolving vacuum energy density (cosmological constant) introduced in the framework of superstring quantum gravity \cite{Ell-Mav-Nan:2000:GENRG1:}. On the other hand, the present accelerated state of the Hubble flow can be explained in the framework of the Tolman-Bondi models \cite{Wil:2007:PHYRL:} as a consequence of inhomogeneous distribution of matter in space. 

Recent observational data indicate that the allowed equation of state for the dark energy is very close to the case of a repulsive cosmological constant $\Lambda>0$ with $\rho_{\Lambda}\doteq0.745\rho_{\rm crit}$. Therefore, it is relevant to consider the influence of $\Lambda>0$ in cosmological and astrophysical phenomena. The cosmological relevance of $\Lambda>0$ is discussed in the standard textbooks \cite{Mis-Tho-Whe:1973:Gra:,Inv:1997:IntroEinsteinRel:}.

We have studied possible effects of $\Lambda>0$ in astrophysically motivated problems \cite{Stu:2002:JB60:}, concentrating on radial geodesics \cite{Stu:1983:BULAI:} and their implications in cosmological models \cite{Stu:1984:BULAI:,Stu-Cal:1984:ASTSS1:,Stu-Sch:2006:AECIC:}, equilibrium (static) positions of spinning test particles \cite{Stu:1999:ACTPS2:,Stu-Kov:2006:CLAQG:,Stu-Hle:2001:PHYSR4:}, and astrophysically most important disc accretion \cite{Stu:2005:MODPLA:}, investigating circular orbits of test particles in the equatorial plane \cite{Stu-Hle:1999:PHYSR4:,Stu-Sla:2004:PHYSR4:} and equilibrium configurations of barotropic perfect fluid tori in black-hole (BH) and naked-singularity (NS) Schwarzschild--de Sitter (SdS) and Kerr--de Sitter (KdS) spacetimes \cite{Stu-Sla-Hle:2000:ASTRA:,Sla-Stu:2005:CLAQG:}. Elliptic integrals of  more general motion of test particles are described in \cite{Kra:2004:CLAQG:,Kra:2005:CLAQG:,Kra:2007:CLAQG:}. The velocity profiles of the geodetical discs and perfect fluid tori in the locally non-rotating frames were studied in \cite{Asch-etal:2004:ASTRA:,Asch:2008:CHIJAA:,Stu-Sla-Tor-Abr:2005:PHYSR4:,Stu-Sla-Tor:2007a:ASTRA:,Mul-Asch:2007:CLAQG:,Sla-Stu:2007a:CLAQG:}. Further, the test particle and perfect fluid properties were treated also in the framework of the optical reference geometry \cite{Kov-Stu:2006:INTJM:,Kov-Stu:2007:CLAQG:,Stu:1990:BULAI:} allowing introduction of inertial forces in the intuitive Newtonian way \cite{Abr-Car-Las:1988:GENRG2:}. Motion of photons in the Reissner--Nordstr\"{o}m--de Sitter and more general Kerr--Newman--(anti-)de Sitter spacetimes is discussed in the papers \cite{Stu-Cal:1991:GENRG2:,Stu-Hle:2000:CLAQG:}. The optical phenomena were treated, e.g., in \cite{Ser:2008:PHYSR4:,Ser:2009:PHYSR4:,Mul:2008:GENRG2:,Sch-Zai:2008:ASTRA:,Khr-Pom:2008:INTJMD:,Stu-Pls:2004:RAGtime4and5:,Bak-etal:2007:CEJP:,Stu-Hle:2002:ACTPS2:,Vir:2009:PHYSR4:}. 

Self-gravitating fluid solutions of the Tolman-Oppenheimer-Volkoff equations with a non-zero cosmological constant ($\Lambda$TOV) were first studied in \cite{Stu:2000:ACTPS2:} for spherically symmetric stellar-like configurations with uniform energy-density distribution and generalized for cosmological solutions in \cite{Boh:2004:GENRG2:}. Solutions of the $\Lambda$TOV equation for polytropic and adiabatic self-gravitating configurations were investigated in \cite{Hle-Stu-Mra:2004:RAGtime4and5:}. Stability of such stellar-like solutions was discussed in \cite{Stu-Hle:2005:RAGtime6and7:} and \cite{Boh-Har:2005:PHYSR4:}. The self-gravitating spherically symmetric polytropic configurations with the dark energy in the form of quintessence give interesting variety of spacetime singularities \cite{Noj-Odi:2009:PHYLB:} that enlarges the singularities occuring in the models based on the $\Lambda$TOV equation. Here we restrict our attention to the test fluid configurations representing physical properties of black hole backgrounds with the cosmological constant. To some extend our results could reflect also the properties of black hole solutions of the $f(R)$ gravity (for review see, e.g., \cite{Noj-Odi:2008::}) where the cosmological constant term is present. Those solutions generally include other parameters whose influence is worth to study.

The influence of $\Lambda>0$ on the BH-spacetime structure can be properly represented by the dimensionless cosmological parameter $y=\Lambda M^2/3$.\footnote{We use the geometric system of units $(c=G=1)$ hereafter.} For SdS black holes admitting existence of stable circular geodesics, which are necessary for the existence of accretion discs, the cosmological parameter $y<y_{\rm ms,e}\doteq 0.000237$ \cite{Stu-Hle:1999:PHYSR4:}. Cosmological tests using the supernova magnitude-redshift relation and the measurements of CMB fluctuations \cite{Spe-etal:2003:ASTJS:,Spe-etal:2007:ASTJS:} imply $\Lambda_0\sim 10^{-56}$cm$^{-2}$, and thus very low present values of $y$ for astrophysically realistic black holes. The most massive currently observed black hole seems to be harboured in the quasar TON~618, having the mass of $6.6\times 10^{10}$\,M$_{\odot}$ \cite{She-etal:2004:ASTRJ2:,Zio:2008:CHIJAA:}. In this case, $y\doteq 4.1\times 10^{-25}$. For primordial black holes in the very early universe, with expected high values of the effective cosmological constant, the values of $y$ can be much closer to $y_{\rm ms,e}$. Considering the electroweak phase transition at $T_{\rm ew}\sim 100$\,GeV, we obtain an estimate of the primordial effective cosmological constant $\Lambda_{\rm ew}\doteq 0.028$\,cm$^{-2}$, while at the level of the quark confinement at $T_{\rm qc}\sim 1$\,GeV we obtain $\Lambda_{\rm qc}\sim 2.8\times10^{-10}$\,cm$^{-2}$ and consequently higher values of $y$ \cite{Stu:2005:MODPLA:,Stu-Sla-Hle:2000:ASTRA:}. 

Geometrically thin (Keplerian) discs are characterized by the quasi-circular motion of test particles along stable circular orbits (geodesics) in the equatorial plane of a~given spacetime.  The inner edge of the Keplerian disc thus corresponds to the innermost stable circular orbit. In the SdS and KdS backgrounds, there is also the outer marginally stable circular geodesic which puts a~natural limit on the maximal extension of Keplerian discs in the spacetimes with $\Lambda>0$ \cite{Stu:2005:MODPLA:}. 

Equilibrium configurations of barotropic perfect fluid tori are described by equipotential surfaces of the `gravito-centrifugal' potential $W$, which correspond to the surfaces of constant pressure in the fluid. Analyzing properties of the equipotential surfaces in the case of marginally stable tori, i.e. tori with uniform distribution of the specific angular momentum $\ell(r,\,\theta)=\mbox{const}$, orbiting in the SdS and KdS backgrounds, three astrophysically relevant phenomena were found \cite{Stu-Sla-Hle:2000:ASTRA:,Sla-Stu:2005:CLAQG:}: (i) the outer cusp in the structure of closed equipotential surfaces, corresponding to the outermost edge of the torus, (ii) strong collimation of open equipotential surfaces near the axis of rotation of the disc that could have some relevance for the collimation of huge jets outside some giant active galaxies, (iii) for current value of the effective cosmological constant $\Lambda_0\sim 10^{-56}$\,cm$^{-2}$, there is a~dimensional coincidence between extension of large galaxies and maximal extension of perfect fluid tori orbiting supermassive black holes with mass in the range of ($10^{6}$--$10^{10}$)\,M$_{\odot}$. For this interval of masses and current cosmological constant $\Lambda_0$, the cosmological parameter $y$ is in the range ($10^{-34}$--$10^{-26}$). Important study of stability of fluid tori in the SdS spacetimes against the so-called runaway instability was realized in \cite{Rez-Zan-Fon:2003:ASTRA:}, suggesting a strong stabilizing effect of the outflow of matter through the outer cusp of toroidal thick discs. Therefore, the cosmological constant strongly influences the structure of disc configurations (both the geometrically thin and thick) introducing quite naturally the outer edge of accretion discs.

It should be stressed that all of the relevant effects of $\Lambda>0$ on the BH physics are well expressed in the SdS spacetime, since the effects of dragging of inertial frames caused by the BH rotation are concentrated in the vicinity of the BH horizon, where the influence of $\Lambda>0$ can be abandoned for realistic values of the BH mass and the present relict cosmological constant $\Lambda_0$. Of course, the efficiency $\eta$ of the accretion process is influenced by the frame dragging in the innermost part of the disc, where the KdS spacetime structure is relevant. For near-extreme KdS black holes characterized by the cosmological parameter $y\ll 10^{-6}$, $\eta\doteq 0.4$, which is much higher in comparison with the SdS black holes where $\eta\doteq 0.059$. Therefore, in studying the large scale properties of disc structures, investigation of the SdS spacetime is quite sufficient, only the accretion efficiency has to be given by the KdS spacetime structure.

One of important problems that could be hardly solved in the framework of full general relativistic (GR) approach is the influence of $\Lambda>0$ on the structure of self-graviting discs. For the current stellar-mass and super-massive SdS black holes, the strong gravity near the BH horizon $r_{\rm h}\approx 2M$ weakens with the distance from the black hole growing and, at $r\gg M$, the structure of the disc can be well described by the Newtonian theory. However, the Newtonian theory starts to lose its validity behind the so-called static radius $r_{\rm s}=y^{-1/3}M$, where the repulsive effect of the cosmological constant starts to be relevant up to the other strong gravity region near the cosmological horizon $r_{\rm c}\approx y^{-1/2}M$. We expect that the pseudo-Newtonian (PN) approach, based on the replacing of the Newtonian gravitational potential by an appropriate PN gravitational potential in the framework of the Newtonian theory, could be succesfull in order to describe properly some GR effects. The PN approach enables to use directly the standard techniques developed in the framework of the Newtonian physics, e.g. the Newtonian discoseismology \cite{Kat-Fuk-Min:1998:BHAccDis:}, and include the repulsive cosmological constant and some other relativistic effects into account. 

These are the reasons that led us to introduce, in close analogy with the standard approach of Paczy\'nski and Wiita \cite{Pac-Wii:1980:ASTRA:} related to the Schwarzschild spacetime, a~PN gravitational potential for the SdS spacetime, and test its accuracy by comparing its predictions with the exact GR results \cite{Stu-Kov:2008:INTJMD:}. We focused on the properties which play a~crucial role for a~geodesic motion and accretion processes in thin accretion discs, e.g. locations of the marginally stable and marginally bound circular geodesics. We concluded that the  PN gravitational potential introduced in \cite{Stu-Kov:2008:INTJMD:} seems to be suitable for the thin disc description in astrophysically realistic SdS spacetimes with the cosmological parameter $y<10^{-25}$ \cite{Stu-Kov:2008:INTJMD:}. 

In the present paper, we study the PN description of perfect fluid tori orbiting the SdS black holes that approximate realistic thick accretion discs, and compare the PN results with the exact GR results. The influence of the discs on the spacetime geometry is assumed to be negligible. We study both the global geometrical structure of toroidal configurations and internal physical properties; the latter under simplified assumption of adiabatic tori. In section \ref{s2}, we introduce the PN gravitational potential for the SdS spacetime and briefly summarize its significance for geodesical motion. In section \ref{s3}, equilibrium configurations of barotropic perfect fluid tori are calculated using both the exact GR and approximate PN approaches. Namely, we  calculate loci of cusps of the equilibrium configurations, shapes of the barotropic tori and depths of the related potential wells; the GR and PN results are compared. In section \ref{s4}, we study properties of the adiabatic tori in both the approaches, i.e., we determine mass-density, pressure and temperature profiles. In section \ref{s5}, we calculate total mass of adiabatic tori in both the approaches and give some limits on the test-disc approximation used in our calculations. In section \ref{dis}, behaviour of thermodynamic quantities across the adiabatic tori and the definition of the PN potential are discussed. In section \ref{con}, some concluding remarks are presented.

\section{Pseudo-Newtonian gravitational and effective potentials}
\label{s2}
According to a general heuristic method \cite{Muk:2002:ASTRJ2:}, 
the PN gravitational potential can be defined by the relation
\begin{equation}                                                                \label{e1}
\psi=\int\frac{\ell_{\rm c}^2}{r^3}\d r,
\end{equation}
where $r$ represents the radial coordinate and $\ell_{\rm c}=L_{\rm c}/E_{\rm c}$ is the GR specific angular momentum, i.e., the ratio of the conserved angular momentum and energy per particle mass, related to circular geodesics in the equatorial (central) plane. This definition is based on the Newtonian relation for the gravitational potential $\psi_{\rm N}=\int l_{\rm c}^2/r^3 \d r$, where $l_{\rm c}$ is the Newtonian angular momentum per mass of the particle moving along the circular orbit.\footnote{The PN potential given by Eq. (\ref{e1}) is defined to match the Newtonian angular momentum per particle mass on a~circular orbit with the GR specific angular momentum. We discuss this kind of the PN potential definition in section \ref{dis}.} As shown below, this definition of the PN gravitational potential yields a number of advantages. Note that it also implies the well-known Paczy\'nski-Wiita gravitational potential \cite{Pac-Wii:1980:ASTRA:} 
\be                                                                            \label{e1a}
\psi_{\rm PW}=-\frac{M}{r-2M},
\ee
describing with high precision the test particle motion and accretion disc structure in the field of any Schwarzschild black hole.

\subsection{Circular geodesics in SdS spacetimes} 
\begin{figure}
\centering
\includegraphics[width=0.75\hsize]{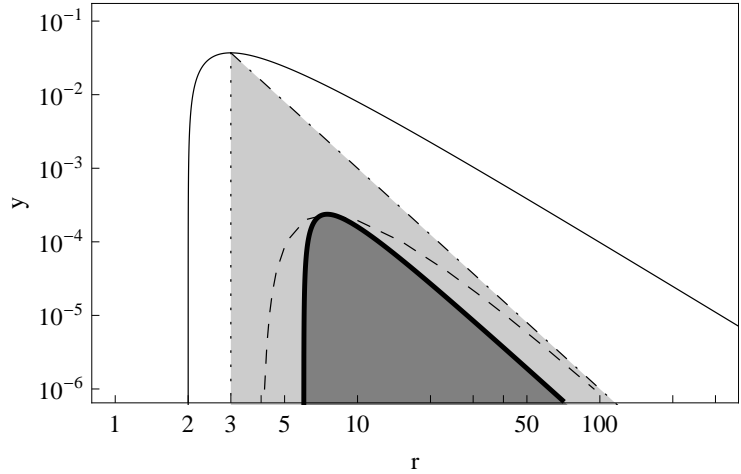}
\caption{Characteristic radii of the SdS spacetime in dependence of the cosmological parameter $y$. Solid curve determines the BH and cosmological horizons, dashed-dotted line determines the static radius, dotted line shows location of the photon circular orbit at $r=3$, dashed curve determines marginally bound circular orbits and the full thick curve marginally stable circular orbits. Circular orbits exist in the shaded region only, its dark part corresponds to the stable ones. Black-hole spacetimes exist for $y<1/27 \doteq 0.037$, those admitting motion along stable circular orbits exist for $y<12/15^4 \doteq 0.000237$.}
\label{f1}
\end{figure}

Geometry of the SdS spacetime in Schwarzschild coordinates $(t,\,r,\,\theta,\,\phi)$ and geometrical units $(c=G=1)$ is described by the line element
\begin{eqnarray}                                                               \label{e2}
\d s^2 = &-& \left(1-\frac{2M}{r}-yr^2\right){\rm d}t^2 \\ 
&+& \left(1-\frac{2M}{r}-yr^2\right)^{-1}{\rm d}r^2+r^2({\rm d}\theta^2+ \sin^2{\theta}\,\d\phi^2),                                                                     \nonumber
\end{eqnarray}
where $M$ is the total mass of the central object (e.g. a~black hole). In the following, we put $M=1$ in order to have completely dimensionless formulae. 
Pseudo-singularities of the line element (\ref{e2}) give the loci of the BH and cosmological horizons
\begin{equation}                                                               \label{e3}
r_{\rm h}=\frac{2}{\sqrt{3y}}\cos{\frac{\pi+\xi}{3}},\quad
r_{\rm c}=\frac{2}{\sqrt{3y}}\cos{\frac{\pi-\xi}{3}},
\end{equation}
where $\xi=\cos^{-1}{(3\sqrt{3y})}$. Both horizons are determined by the condition 
\begin{equation}                                                               \label{e4}
y=y_{\rm h}(r)\equiv\frac{r-2}{r^3},
\end{equation}
and exist for $0<y<y_{\rm crit}=1/27$, separating the spacetime into two dynamic regions and one static region. For $y=y_{\rm crit}$, both the horizons coalesce at the photon circular orbit located at $r=3$. For $y>1/27$, the horizons disappear and the corresponding SdS spacetimes become dynamic NS spacetimes (figure \ref{f1}). 

The geodesic motion in the SdS spacetime is fixed to central planes. In the equatorial plane ($\theta=\pi/2$), the motion is determined by the GR effective potential \cite{Stu-Hle:1999:PHYSR4:}
\begin{equation}                                                              \label{e5}
V^2_{\rm eff}(r;\,y,\,L)=\left(1-\frac{2}{r}-yr^2\right)\left(1+\frac{L^2}{r^2}\right),
\end{equation}
where $L$ is the angular momentum per unit rest mass of the particle. The motion is allowed in the region where particle's energy per unit rest mass $E\geq V_{\rm eff}$.
Circular geodesics are determined by local extrema of the effective potential. At a given radius they are characterized by the constants of motion 
\begin{eqnarray}                                                              \label{e6}
E_{\rm c}&=&\left(1-\frac{2}{r}-yr^2\right)\left(1-\frac{3}{r}\right)^{-1/2}, \\
L_{\rm c}&=&\left[r(1-yr^3)\right]^{1/2}\left(1-\frac{3}{r}\right)^{-1/2}.    \label{e7}
\end{eqnarray}
Corresponding GR specific angular momentum then takes the form
\begin{equation}                                                              \label{e8}
\ell_{\rm c}=\frac{[r^3(1-yr^3)]^{1/2}}{(r-2-yr^3)}.
\end{equation}

Circular orbits of test particles exist in the region
\be                                                                           \label{e8a}
3<r\leq r_{\rm s}\equiv y^{-1/3},
\ee
where the lower limit is given by the photon circular orbit, which is located at $r=3$ independently of the cosmological parameter $y$, see \cite{Stu-Hle:1999:PHYSR4:} for more details, and the upper limit corresponds to the so-called \emph{static radius}, which is the place where a~geodesic observer with just time-component of its 4-momentum being non-zero can exist. At the static radius, the gravitational attraction of the black hole is just balanced by the cosmic repulsion represented by the cosmological constant, i.e., test particles with zero angular momentum feel no `force' there, similarly to the radial infinity in asymptotically flat spacetimes.

\subsection{Pseudo-Newtonian gravitational potential}
Constructing the PN gravitational potential for the SdS spacetime, we have to reflect appropriately both the gravitational attraction of the black hole and the repulsive effects of the cosmological constant. The PN gravitational potential is defined by relation (\ref{e1}) in which the specific angular momentum of a particle on circular orbit is given by relation (\ref{e8}). After integration, we obtain a general form of the potential $\psi$, i.e.
\be                                                                           \label{e9a}
\psi=-\frac{r}{2(r-2-yr^3)}+\mathcal{C},
\ee
where $\mathcal{C}$ is an integration constant having no physical meaning but enabling to specify a proper form of the potential $\psi$. Here, we demand that for $y=0$ expression (\ref{e9a}) takes the form of Paczy\'nski-Wiita potential (\ref{e1a}), which gives $\mathcal{C}=1/2$. The PN gravitational potential for the SdS spacetime then has the particular simple form\footnote{Another possibility is to demand that at the static radius, where the potential (\ref{e9a}) has a~local maximum, $\psi(r_{\rm s})=0$ in analogy with the asymptotic behaviour of the Paczy\'nski-Wiita potential in the asymptotically flat Schwarzschild spacetime. This gives $\mathcal{C}=1/2(1-3y^{1/3})$ implying the PN potential in the form
\begin{eqnarray*}
\psi=\frac{r^3y-3ry^{1/3}+2}{2(3y^{1/3}-1)(r-2-r^3y)}.
\end{eqnarray*}
Such a choice is used in \cite{Stu-Kov:2008:INTJMD:}. Both particular formulae of the potential, however, describe the same physics.}
\be                                                                           \label{e9b}
\psi=-\frac{1+yr^{3}/2}{r-2-yr^3}.
\ee

\begin{figure}
\centering
\includegraphics[width=0.75\hsize]{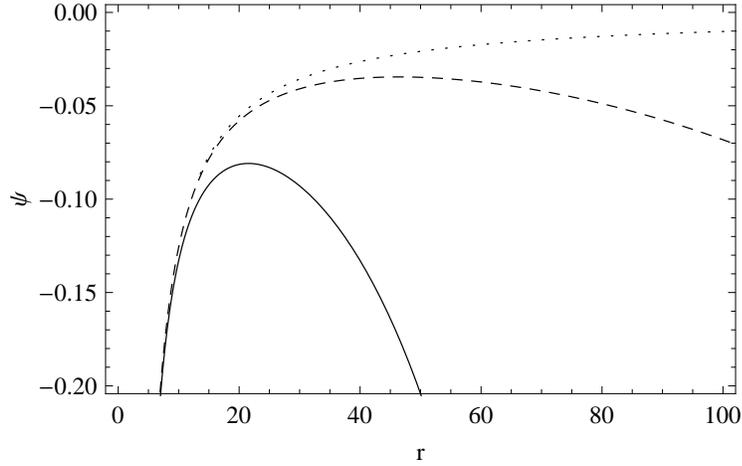}
\caption{PN gravitational potential $\psi$ given by relation (\ref{e9b}) for three values of the cosmological parameter $y$: $10^{-4}$ (solid), $10^{-5}$ (dashed), and $0$ (dotted).}
\label{f2}
\end{figure}

The PN gravitational potential $\psi$ diverges at horizons of the SdS spacetime, and reaches its maximal value at the static radius of the spacetime where the gradient $\nabla\psi$ changes its sign from positive to negative values, reflecting both the positions of black-hole and cosmological horizons and the changing character of the gravitational field which becomes repulsive for $r>r_{\rm s}$ (figure~\ref{f2}).
 
As standard practise in mechanics, we can construct the PN effective potential for the radial motion by the relation  
\begin{equation}                                                              \label{e10}
v_{\rm eff}(r,\theta;y,l)=\psi(r;y)+\frac{l^2}{2r^2\sin^2\theta},
\end{equation}
where $l$ is the PN angular momentum per particle mass. In the PN approach, as well as in the Newtonian theory, the motion of a~test particle is confined to a central plane, which is conveniently chosen to be the equatorial plane ($\theta=\pi/2$). Circular orbits (in the equatorial plane) are again given by local extrema of the effective potential $v_{\rm eff}$ (\ref{e10}) with $\theta=\pi/2$. Related energy and angular momentum per particle mass are given by the relations
\begin{eqnarray}                                                              \label{e11a}
e_{\rm c}=\frac{1}{2}\left[ 1-\frac{r(r-3)}{(r-2-yr^3)^2}\right], \\
l_{\rm c}=\ell_{\rm c}=\frac{[r^3(1-yr^3)]^{1/2}}{r-2-yr^3}.                   \label{e11b}
\end{eqnarray}

\subsection{Comparison of general relativistic and pseudo-Newtonian effective potentials}
Despite of the fact that the PN effective potential $v_{\rm eff}$ differs from its GR counterpart $V_{\rm eff}$, it enables to determine basic features of the test particle motion in accord with GR results, especially in the case of the circular orbits.
Since behaviour of the angular momenta $L_{\rm c}$ and $l_{\rm c}$, governing local extrema of the effective potentials $V_{\rm eff}$ and $v_{\rm eff}$, is similar, and positions of the $L_{\rm c}$ and $l_{\rm c}$ local extrema coincide, the PN calculation of the circular geodesics loci corresponds to the GR results. Marginally stable circular geodesics are determined by the condition
\begin{equation}                                                              \label{e12}
y=y_{\rm ms}(r)\equiv \frac{r-6}{r^3(4r-15)}
\end{equation}
that is identical with the GR relation \cite{Stu-Hle:1999:PHYSR4:}. Maximum of the function $y_{\rm ms}(r)$ takes the value $y_{\rm ms,e}=12/15^4$, which represents the limiting value for the SdS spacetimes admitting stable circular geodesics and, therefore, for the existence of accretion discs as well \cite{Stu-Sla-Hle:2000:ASTRA:} (figure~\ref{f1}).

The PN energy per unit mass $e_{\rm c}$ and its GR counterpart $E_{\rm c}$ are related by the formula
\begin{equation}                                                              \label{e13}
e_{\rm c}=\frac{1}{2}\left(1-\frac{1}{E_{\rm c}^{2}}\right).
\end{equation}
Therefore, loci of the marginally bound circular geodesics, corresponding to two unstable circular geodesics with the same energy $E_{\rm c,mb}$, coincide with two unstable circular orbits with the same energy $e_{\rm c,mb}$ and angular momentum $l_{\rm c,mb}=L_{\rm c,mb}/E_{\rm c,mb}$ obtained by the PN approach.

A detailed comparison of the GR and PN approach related to the test particle circular motion is presented in \cite{Stu-Kov:2008:INTJMD:} and will not be repeated here. We only summarize that for $y<10^{-6}$ the PN formalism reflects quite well the properties of test particle motion down to the marginally bound circular orbit with precission growing with the cosmological parameter $y$ descending. On the other hand, the PN gravitational potential does not reflect the existence of the photon circular geodesic and the properties of unstable circular geodesics under the inner marginally bound circular geodesic. Nevertheless, these orbits are irrelevant for accretion discs and their properties discussed in the present paper.

\section{Equilibrium configurations of barotropic perfect fluid tori}
\label{s3}
In studies of thick accretion discs with non-negligible pressure gradients, investigation of equilibrium configurations of perfect fluid tori plays a significant role \cite{Fra-Kin-Rai:2002:AccretionPower:}. The tori are considered to be configurations with negligible self-gravity and no influence on the spacetime structure, determined by the BH spacetime geometry and an appropriately chosen rotational law. We assume a barotropic fluid, i.e. a fluid with the pressure-energy density relation of the form $p=p(\epsilon)$, for which, according to Boyer \cite{Boy:1965:PCPS:}, the stationary equilibrium configurations are described by closed equipotential surfaces of the `gravito-centrifugal' potential. Thick discs are characterized by toroidal equipotential surfaces. The necessary condition for the accretion process to be allowed is the existence of critical, marginally closed equipotential surface with an inner cusp, which enables an outflow of matter from the disc into the black hole. In spacetimes with a repulsive cosmological constant ($\Lambda>0$), the critical point (cusp) of the marginally closed equipotential surface can be located also at the outer side, enabling thus the outflow of matter from the torus into the outer space. Such a configuration is called excretion disc. In the special case, both cusps exist  at one critical equipotential surface, enabling simultaneous outflows of matter from the disc into the black hole and to the outer space \cite{Stu-Sla-Hle:2000:ASTRA:,Sla-Stu:2005:CLAQG:}. 

In the following, we show that the PN gravitational potential (\ref{e9b}) enables to construct barotropic perfect fluid tori in the framework of the PN theory\footnote{Within the PN approach, the thick accretion discs orbiting the Schwarzschild black hole were treated in detail, e.g., in \cite{Abr-Cal-Nob:1980:ASTRJ2:}.} with properties very close to relativistic thick discs. 

\subsection{General relativistic perfect fluid tori}
The stress-energy tensor of a~perfect fluid with an~energy density $\epsilon$ and a~pressure $p$ (as measured by comoving observers) orbiting a black hole is given by the relation $T^{ik}=(\epsilon+p)U^i U^k+pg^{ik}$. The fluid is assumed to move in azimuthal direction only, thus the 4-velocity field of the fluid has only two non-zero components: $U^t(r,\theta)$ and $U^{\phi}(r,\theta)$. Projecting the covariant energy-momentum conservation law $\nabla_k T^{ik}=0$ onto the hypersurface orthogonal to the 4-velocity $U^{i}$, we obtain the relativistic Euler equation in the form \cite{Koz-Jar-Abr:1978:ASTRA:,Abr-Jar-Sik:1978:ASTRA:}
\begin{equation}                                                               \label{e14}
\frac{\partial_i p}{p+\epsilon}=-\partial_i(\ln U_t)+\frac{\Omega\partial_i\ell}{1-\Omega\ell}, \quad i=\{r,\theta\}
\end{equation}
where in general stationary and axisymmetric spacetime
\begin{equation}                                                               \label{e15}
(U_t)^2=\frac{g_{t\phi}^2-g_{tt}g_{\phi\phi}}{\ell^2 g_{tt}+2\ell g_{t\phi}+g_{\phi\phi}}.
\end{equation}
The angular velocity (related to distant observers) $\Omega=U^{\phi}/U^t$ and the specific angular momentum $\ell=L/E=-U_{\phi}/U_t$ are related by the formula
\begin{equation}                                                               \label{e16}
\Omega=-\frac{\ell g_{tt}+g_{t\phi}}{\ell g_{t\phi}+g_{\phi\phi}}.
\end{equation}
In the SdS spacetimes, relations (\ref{e15}) and (\ref{e16}) reduce to the form
\begin{equation}                                                               \label{e17}
(U_t)^2=-\frac{g_{tt}g_{\phi\phi}}{\ell^2 g_{tt}+g_{\phi\phi}},\quad \frac{\Omega}{\ell}=-\frac{g_{tt}}{g_{\phi\phi}}.
\end{equation}

For any barotropic fluid, i.e., the fluid for which $p=p(\epsilon)$, a~solution of the relativistic Euler equation can be given in terms of Boyer's condition for constant-pressure surfaces, which coincide with equipotential surfaces of the potential $W(r,\theta)$ defined by the relations \cite{Koz-Jar-Abr:1978:ASTRA:,Abr-Jar-Sik:1978:ASTRA:}
\begin{eqnarray}                                                               \label{e18}
\lefteqn{\int_0^p \frac{{\rm d}p}{p+\epsilon}=W_{\rm in}-W,} \\                \label{e19}
& & W_{\rm in}-W=\ln{(U_t)}_{\rm in}-\ln{(U_t)}+\int_{\ell_{\rm in}}^{\ell}\frac{\Omega\d\ell}{1-\Omega\ell};
\end{eqnarray}
the subscript `in' refers to the inner edge of the disc in the equatorial plane. 
The equipotential surfaces are determined by the condition $W(r,\theta)={\rm const}$ and in a given spacetime can be found from the equation (\ref{e19}), if a rotational law $\Omega=\Omega(\ell)$ is given.
 
In the case of discs with uniform distribution of the specific angular momentum $\ell(r,\theta)=\mbox{const}$\footnote{This case corresponds to the marginally stable tori with properties reflecting quite well those of tori with a general allowed specific angular momentum profile.}, the potential $W$ is given by a~simple formula 
\begin{equation}                                                               \label{e20}
W(r,\theta)=\ln{U_t}=\ln\frac{(r-2-yr^3)^{1/2}r\sin\theta}{[r^3\sin^2\theta-(r-2-yr^3)\ell^2]^{1/2}}.
\end{equation} 
The first reality condition, $r-2-yr^3\geq0$, restricts the existence of equipotential surfaces to stationary regions of the spacetime. The second reality condition, $r^3\sin^2\theta-(r-2-yr^3)\ell^2 >0$, evaluated in the equatorial plane ($\theta=\pi/2$), implies a restriction given by the photon motion  
\begin{equation}                                                               \label{e21}
\ell^2 < \ell^2_{\rm ph}(r;y)\equiv\frac{r^3}{r-2-yr^3},
\end{equation}
as the function $\ell^2_{\rm ph}(r;y)$ plays the role of the effective potential governing the photon geodesic motion in the equatorial plane \cite{Stu-Hle:1999:PHYSR4:}. The function $\ell^2_{\rm ph}(r;y)$ has one local extreme (minimum), $\ell^2_{\rm ph,c}$, located at $r=3$ independently of the cosmological parameter $y$, and corresponding to the circular photon orbit in the equatorial plane.

Character of the equipotential surfaces is fully determined by the existence of local extrema of the
function $W(r,\theta)$ and their mutual behaviour. Since the orbits with vanishing gradient of $W$, i.e., satisfying conditions $\p_r W(r,\theta)=\p_{\theta}W(r,\theta)=0$, correspond to places with zero pressure gradients, the fluid has to follow the geodesic motion there. As shown in \cite{Stu-Sla-Hle:2000:ASTRA:}, the local extrema of $W$ are located only in the equatorial plane ($\theta=\pi/2$), where the radial profile of the potential is described by the function 
\begin{equation}                                                               \label{e22}
W_{\rm \pi/2}(r)\equiv W(r,\theta=\pi/2). 
\end{equation}
The necessary condition for the local extrema, $\partial_{r}W_{\pi/2}(r)=0$, implies
\begin{equation}                                                               \label{e23}
\ell^2=\ell_{\rm c}^2(r;y)\equiv\frac{r^3(1-yr^3)}{(r-2-yr^3)^2}.
\end{equation}
Since $\ell^2_{\rm c}(r;y)$ corresponds to the specific angular momentum distribution of circular geodesics, the extrema of the potential (corresponding to circular geodesics) fulfill the relation 
\begin{equation}                                                               \label{e24}
W_{\rm ext}\equiv W_{\pi/2}(r;\ell=\ell_{\rm c},y)=\ln{E_c(r;y)},
\end{equation}
where $E_{\rm c}$ is the energy per unit rest mass related to the circular geodesics (\ref{e6}).

\begin{figure}
\centering
\begin{minipage}{.49 \linewidth}
\centering
\includegraphics[width=.99 \hsize]{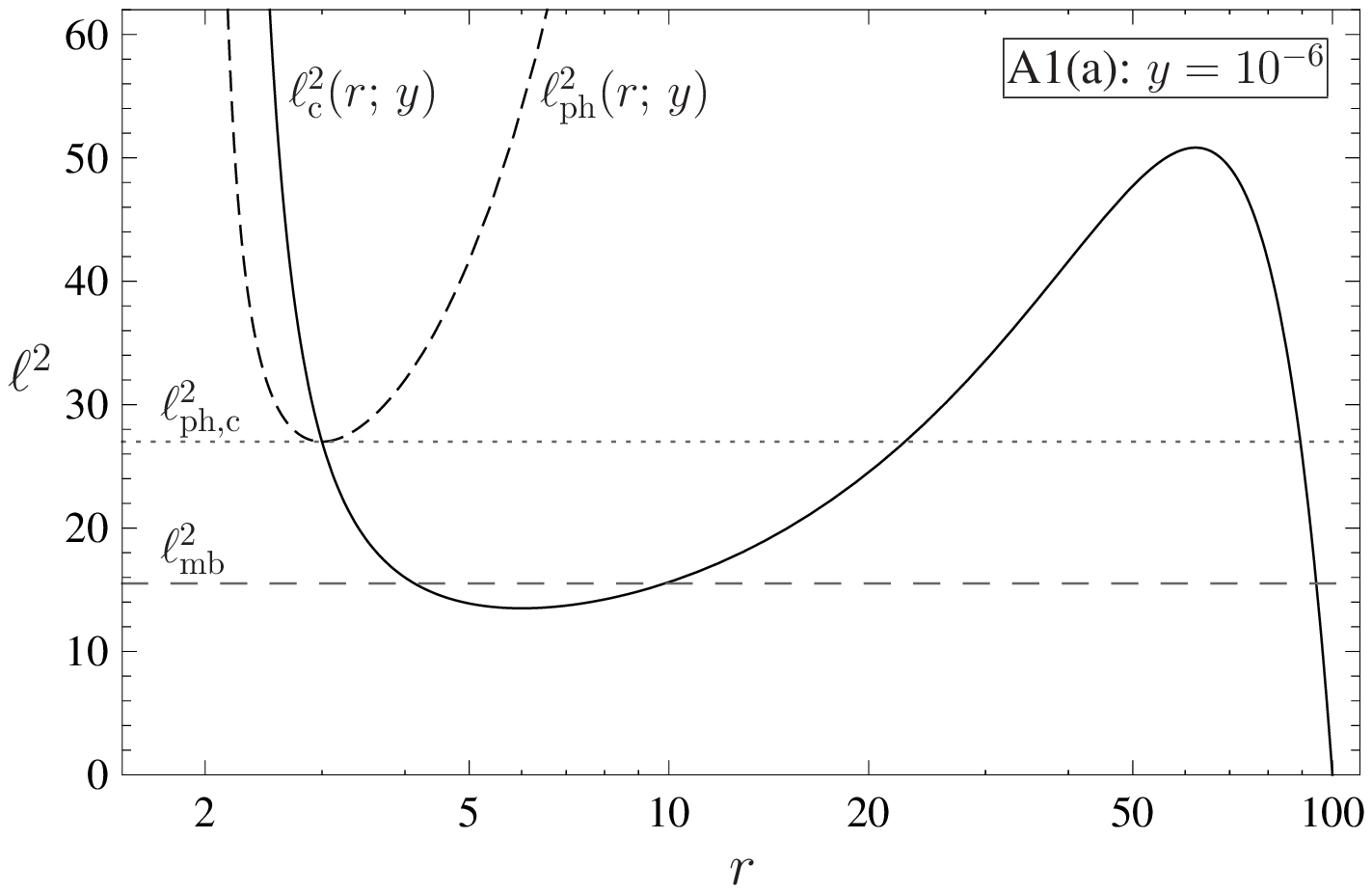}
\end{minipage}\hfill
\begin{minipage}{.49 \linewidth}
\centering
\includegraphics[width=.99 \hsize]{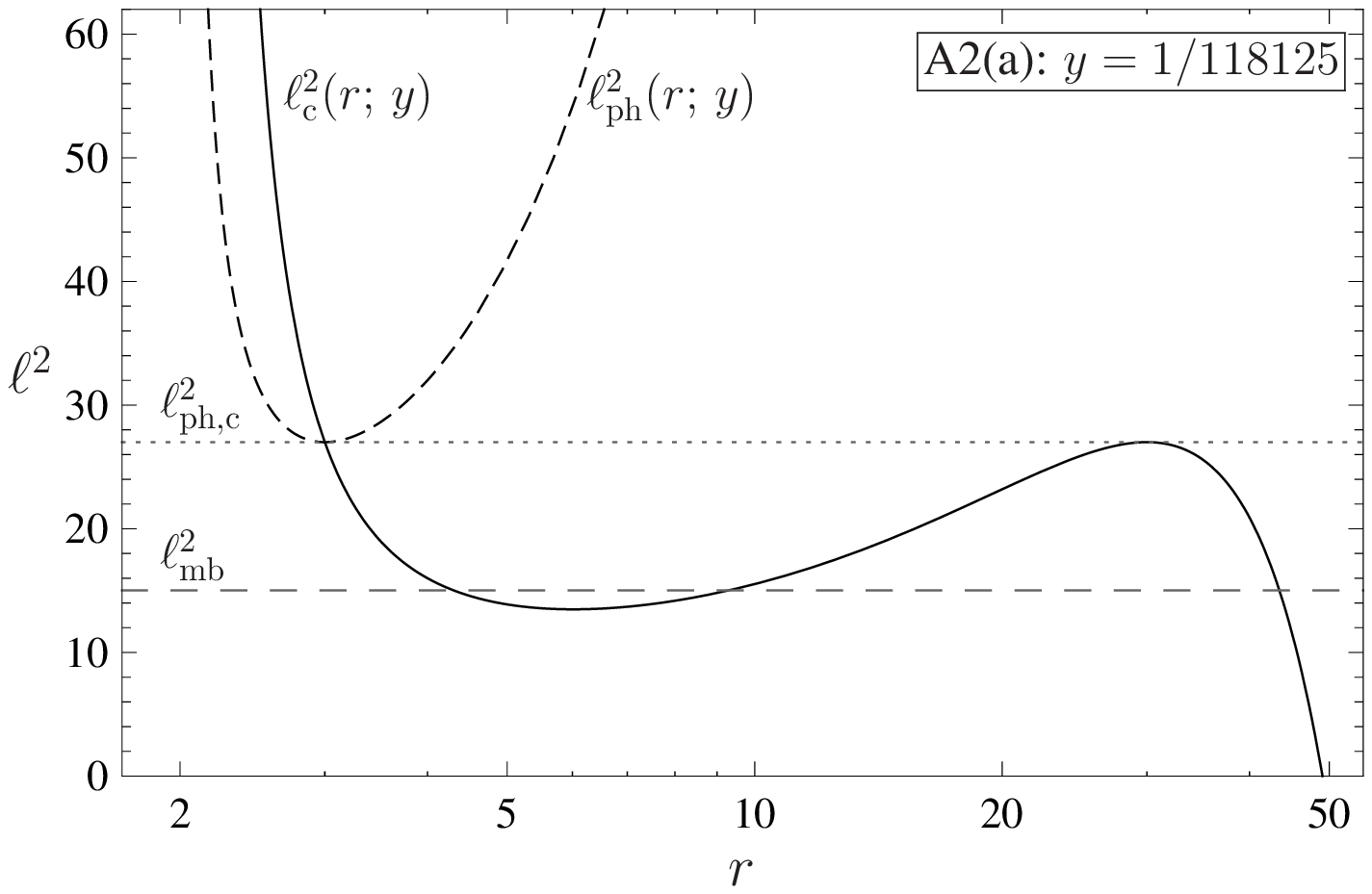}
\end{minipage}
\begin{minipage}{.49 \linewidth}
\centering
\includegraphics[width=.99 \hsize]{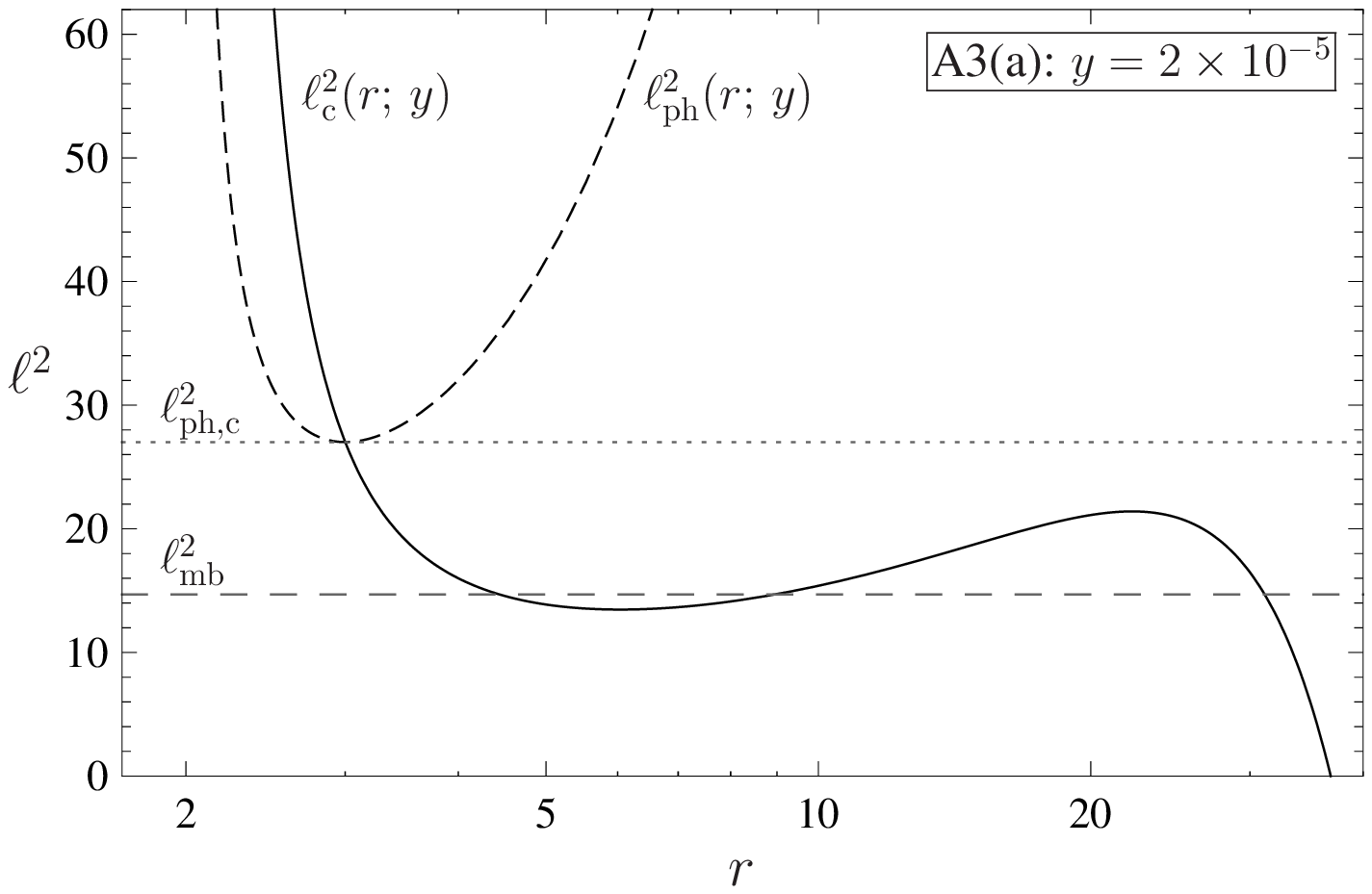}
\end{minipage}\hfill
\begin{minipage}{.49 \linewidth}
\centering
\includegraphics[width=.99 \hsize]{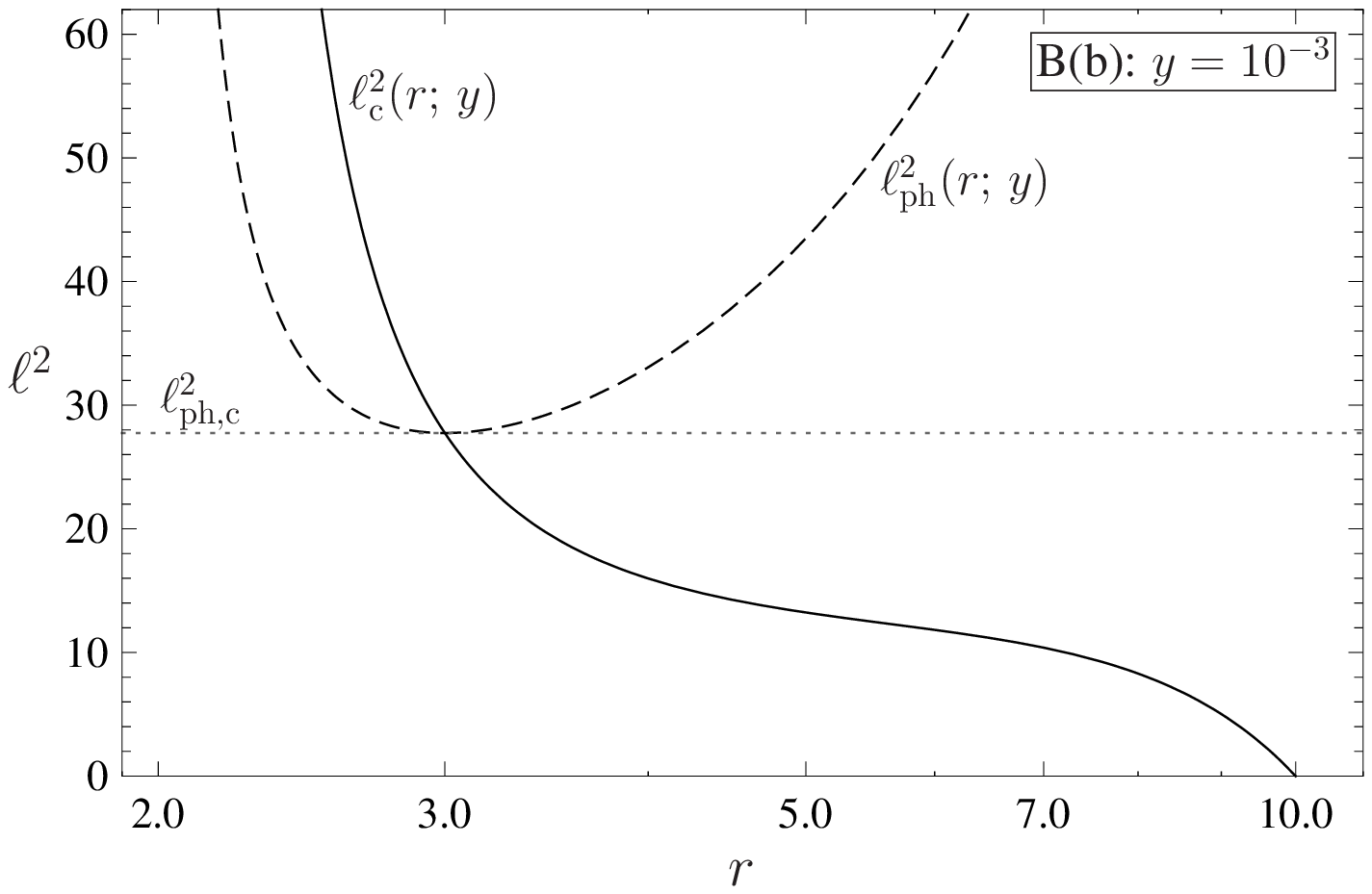}
\end{minipage}
\caption{Functions $\ell^2_{\rm c}(r;y)$ (solid) and $\ell^2_{\rm ph}(r;y)$ (dashed). There are four classes characterizing mutual behaviour of these functions (A1, A2, A3, B) and two different types of behaviour of the function $\ell^2_{\rm c}(r;y)$ (a,b). The horizontal dashed and dotted lines denote values of $\ell^2_{\rm mb}$ and $\ell^2_{\rm ph,c}$ for a given cosmological parameter $y$, respectively.}
\label{f3}
\end{figure}

There are four types of the $\ell_{\rm ph}^2$ and $\ell_{\rm c}^2$ mutual behaviour, depending on the value of the cosmological parameter $y$ (figure~\ref{f3} and table~\ref{t1}). They give four classes of the SdS backgrounds, differing in the types of the structure of equipotential surfaces admitted by the spacetime. For $y_{\rm ms,e}<y<y_{\rm crit}=1/27\doteq 0.037$, the function $\ell_{\rm c}^2(r;y)$ is monotonically decreasing, corresponding to unstable circular geodesics only. For $y<y_{\rm ms,e}=12/15^4 \doteq 2.4\times 10^{-4}$, the function $\ell_{\rm c}^2(r;y)$ has two local extrema, the minimum $\ell^2_{\rm ms,i}$ and the maximum $\ell^2_{\rm ms,o}$, corresponding to the inner and outer marginally stable circular geodesics. The rising part of $\ell_{\rm c}^2(r;y)$, located between the local extrema, corresponds to stable circular geodesics, while the descending parts of $\ell_{\rm c}^2(r;y)$ to unstable ones. If $y=y_{\rm e}=1/118125\doteq 8.5\times 10^{-6}$, $\ell^2_{\rm ms,o}=\ell^2_{\rm ph,c}$, and for $y<y_{\rm e}$, $\ell^2_{\rm ms,o}>\ell^2_{\rm ph,c}$. 

Equilibrium toroidal configurations of a~barotropic fluid can exist only in the SdS spacetimes enabling motion along stable circular geodesics, i.e. the spacetimes with $y<y_{\rm ms,e}$, and with  the specific angular momentum restricted to the range $\ell\in(\ell_{\rm ms,i},\ell_{\rm ms,o})$. For such a constant specific angular momentum distribution, the radii given by the condition $\ell=\ell_{\rm c}(r;y)$ correspond to the motion of the fluid along unstable, stable and again unstable circular geodesic, as the radius grows. The stable circular geodesic represents a center of the torus, because the potential $W(r,\theta)$ has a local minimum there, while the unstable circular geodesics determine critical points (cusps) where the potential $W(r,\theta)$ has local maxima and the corresponding equipotential surface is self-crossing. If the critical surface with the cusp is marginally closed, i.e. it encloses other closed equipotential surfaces corresponding to lower values of the potential $W(r,\theta)$, an outflow of matter from the torus through the cusp is possible. For $\ell\in(\ell_{\rm ms,i},\ell_{\rm mb})$ where $\ell_{\rm mb}$ is the specific angular momentum of a particle moving along the inner/outer marginally bound circular geodesic, the critical marginally closed equipotential surface self-crosses in the inner cusp; this configuration corresponds to the accretion disc. For $\ell\in(\ell_{\rm mb},\ell_{\rm ms,o})$, the critical marginally closed equipotential surface self-crosses in the outer cusp; this configuration corresponds to the excretion disc. For $\ell=\ell_{\rm mb}$, the critical marginally closed equipotential surface possesses both cusps, therefore, common accretion onto the BH and excretion into the outer space are equally possible; this configuration is called `marginally bound accretion disc'. When $y<y_{\rm e}$ and $\ell_{\rm ph,c}<\ell<\ell_{\rm ms,o}$, a forbidden region for the fluid with a given $\ell=\mbox{const}$ exists, restricted by the radii following from the relation  $\ell=\ell_{\rm ph}(r;y)$.

\begin{table}
\caption{Classification of the SdS spacetimes. There are four types of the SdS spacetimes differing in the behaviour of the functions $\ell^2_c(r;y)$ and $\ell^2_{\rm ph}(r;y)$ (relevant for the GR classification) and two types of behaviour of function $l^2_{\rm c}(r;y)$ (relevant for the PN classification). Note that in realistic astrophysical situations the case A1 is relevant.}
\begin{indented}
\item[]\begin{tabular}{@{}lll}
\br
PN & Relativistic & Range of $y$ \\
class & class & \\
\mr
a&A1&$0<y<y_{\rm e}$\\
a&A2&$y=y_{\rm e}$\\
a&A3&$y_{\rm e}<y<y_{\rm ms,e}$\\
b&B&$y_{\rm ms,e}\leq y<y_{\rm crit}$\\
\br
\end{tabular}
\end{indented}
\label{t1}
\end{table}

Due to the properties of the functions $\ell^2_{\rm ph}(r;y)$ and $\ell^2_{\rm c}(r;y)$, we can deduce that there exist twelve characteristic structures of the equipotential surfaces; for the astrophysically most relevant relativistic class A1 (see table \ref{t1}), nine characteristic structures of the equipotential surfaces exist and five of them correspond to toroidal configurations (for details see \cite{Stu-Sla-Hle:2000:ASTRA:}). In dependence on $\ell$, the configuration describes the accretion disc, marginally bound accretion disc, excretion disc without the forbidden region, excretion disc with the forbidden region in the form of infinitesimally thin shell, and the excretion disc with the finite forbidden region.

\subsection{Pseudo-Newtonian perfect fluid tori}
The PN version of the Euler equation for the perfect fluid characterized by the pressure $p$ and the mass-density $\rho$, rotating in a central gravitational field, takes the form \cite{Pac-Wii:1980:ASTRA:}
\begin{equation}                                                              \label{e25}
\frac{1}{\rho}\partial_i p=-\partial_i w, \quad i=\{r,\theta\},
\end{equation}
where the potential $w$ is defined by the relation 
\begin{equation}                                                              \label{e26}
w=\psi-\int\frac{l^2}{R^3}\,\d R,
\end{equation}
and $R=r\sin\theta$. 
Again, for a barotropic fluid, $p=p(\rho)$, the surfaces of constant pressure are just the equipotential surfaces of the potential $w$:
\begin{equation}                                                              \label{e27}
\int_{0}^p {\frac{{\rm d}p}{\rho}}=w_{\rm in}-w.
\end{equation}
In the SdS spacetimes, assuming the PN gravitational potential $\psi$ given by relation (\ref{e9b}) and uniform distribution of the angular momentum per unit mass, $l={\rm const}$, we obtain the potential $w$ in the form\footnote{Constant of integration in (\ref{e26}) is chosen to be zero.}
\begin{equation}                                                              \label{e28}
w(r,\theta) = -\frac{1+yr^{3}/2}{r-2-yr^3} + \frac{l^2}{2r^2\sin^2\theta},
\end{equation}
which is equal to the PN effective potential governing the radial motion (\ref{e10}).

The PN potential $w$ diverges at horizons of the SdS spacetime and is regular in the whole stationary region of the spacetime, but it does not reflect the existence of the photon circular orbit \cite{Stu-Kov:2008:INTJMD:}. 
Its radial profile in the equatorial plane is described by the function
\begin{equation}                                                              \label{e29}
w_{\rm \pi/2}(r)\equiv w(r,\theta=\pi/2).
\end{equation}
Extrema of $w_{\rm \pi/2}(r)$ correspond to the motion along Keplerian circular orbits with the energy per unit mass $e=e_{\rm c}$ and the angular momentum per unit mass $l=l_{\rm c}$ given by (\ref{e11a}) and (\ref{e11b}),
\begin{equation}                                                              \label{e30}
w_{\rm ext}\equiv w_{\rm \pi/2}(r;l=l_{\rm c},y)=e_c(r;y).
\end{equation}

In the case of the PN potential $w$, the structure of its equipotential surfaces is governed by the function $\ell_{\rm c}^2(r;y)$ only, and we can find seven various structures of equipotential surfaces in dependence on the parameters $y$ and $l$. For $y<y_{\rm ms,e}$ and $l\in(\ell_{\rm ms,i},\ell_{\rm ms,o})$, the closed equipotential surfaces of the potential $w$ exist, corresponding to barotropic fluid tori. The center and cusps are given in exactly the same way as in the GR framework. Again, we obtain configurations corresponding to accretion discs, marginally bound accretion discs and excretion discs (figure \ref{f4}).

\begin{figure}
\begin{minipage}{1\linewidth}
\centering
\includegraphics[width=.9 \hsize]{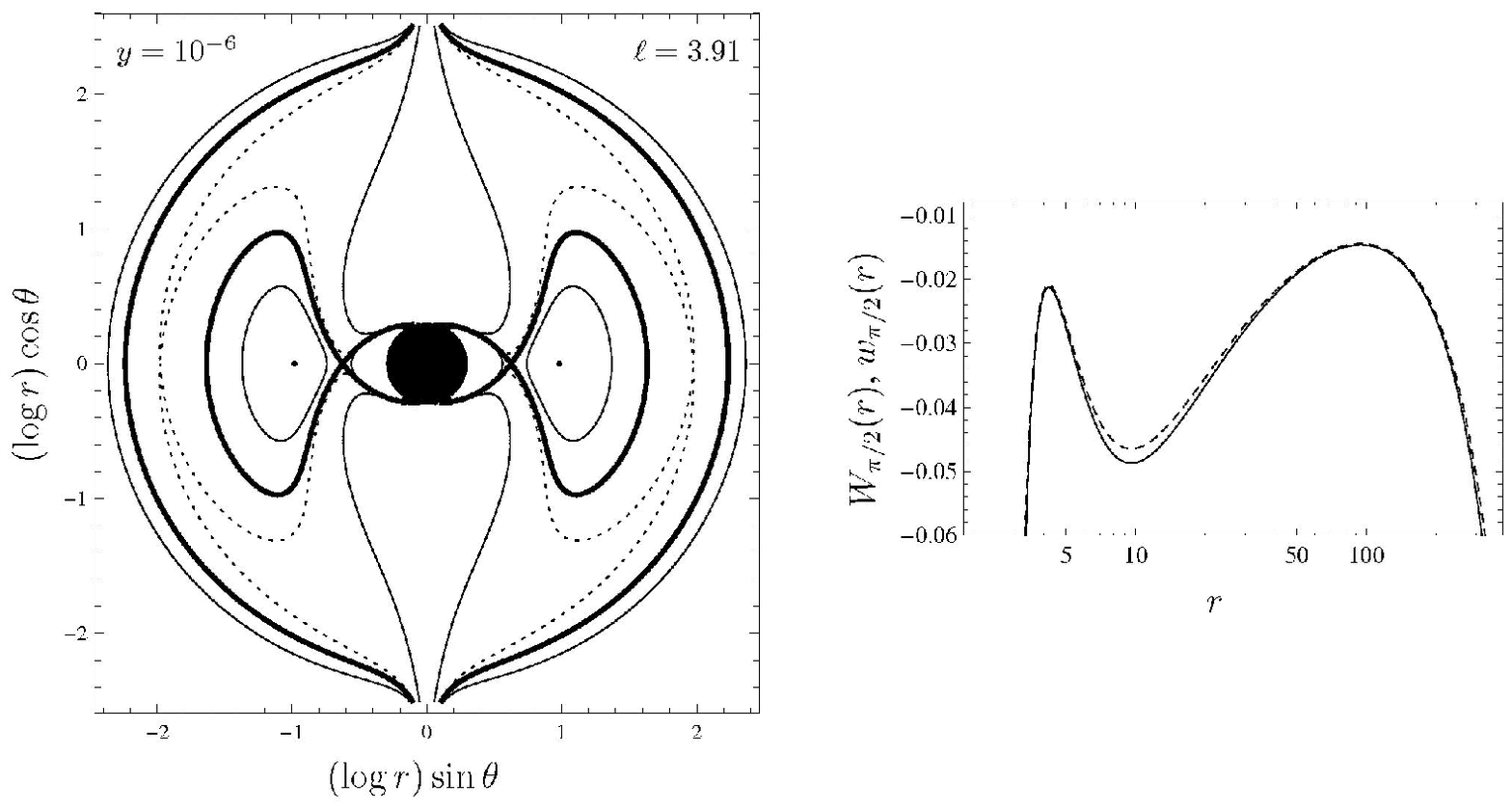}
\par\centering \small(a)\enspace
\end{minipage}
\vskip2ex
\begin{minipage}{1\linewidth}
\centering
\includegraphics[width=.9 \hsize]{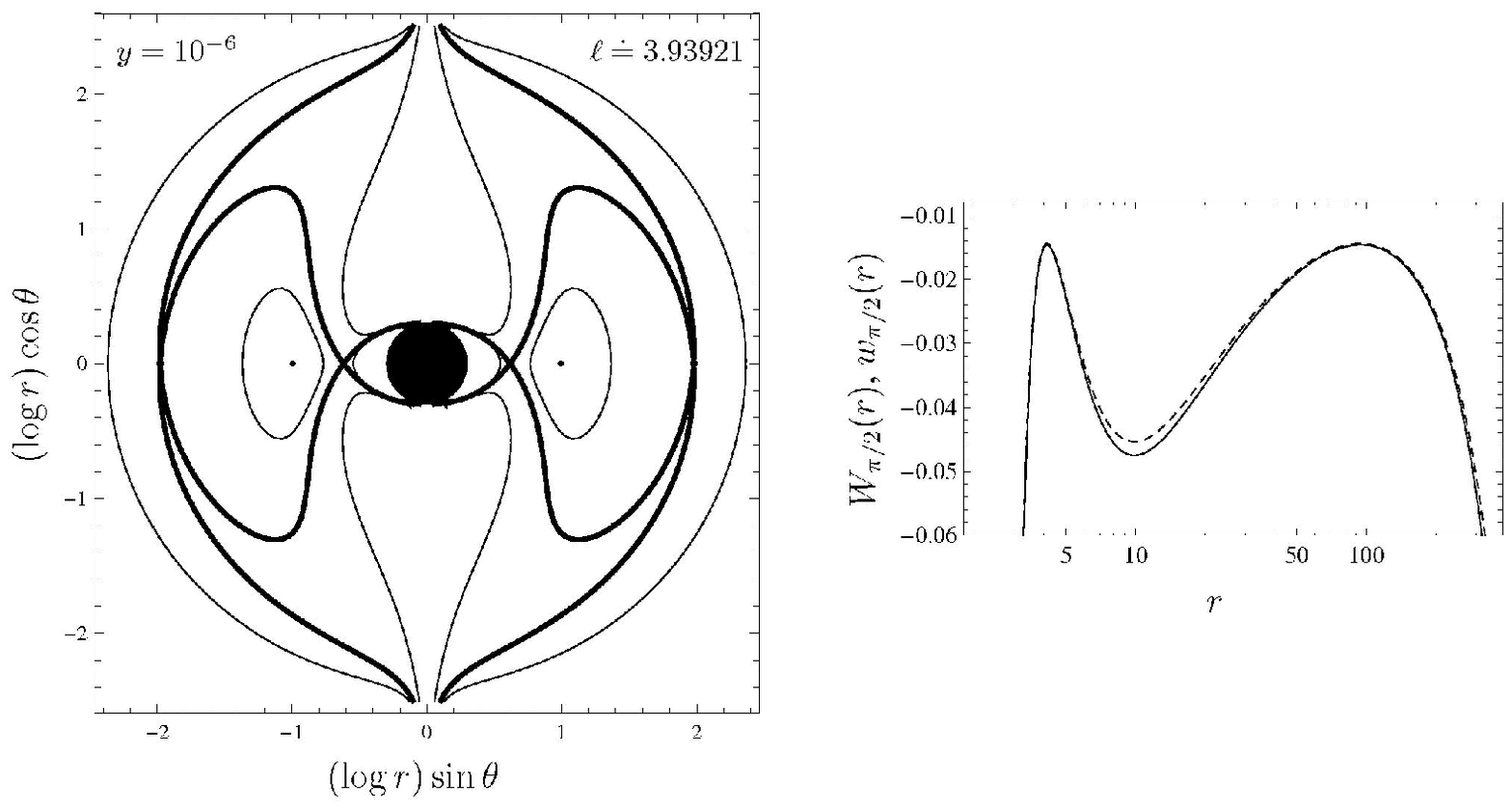}
\par\centering \small(b)\enspace
\end{minipage}
\caption{Meridional sections of equipotential surfaces (potential contours) of the potentials $W(r,\theta)$ and $w(r,\theta)$, and their behaviour in the equatorial plane ($\theta=\pi/2$) represented by the functions $W_{\pi/2}(r)$--dashed curve and $w_{\pi/2}(r)$--solid curve. Self-crossing (critical) equipotential surfaces correspond to the values of potentials at their local maxima. (a) $\ell_{\rm ms,i}<\ell<\ell_{\rm mb}$: the inner critical surface is marginally closed, the outer critical surface is open. (b) $\ell=\ell_{\rm mb}$: the only critical surface containing both cusps is marginally closed. (c) $\ell_{\rm mb}<\ell<\ell_{\rm ph,c}<\ell_{\rm ms,o}$ or $\ell_{\rm mb}<\ell<\ell_{\rm ms,o}<\ell_{\rm ph,c}$: the inner critical surface is open, the outer critical surface is marginally closed. (d) $\ell_{\rm ph,c}<\ell<\ell_{\rm ms,o}$: In GR approach, there is a `forbidden region' for the fluid with prescribed specific angular momentum and, thus, no inner (open) critical surface; the outer critical surface is marginally closed. In PN approach, the `forbidden region' does not exist, the inner critical surface is open, the outer critical surface is marginally closed.}
\label{f4}
\end{figure}
\addtocounter{figure}{-1}
\begin{figure}
\begin{minipage}{1\linewidth}
\centering
\includegraphics[width=.9 \hsize]{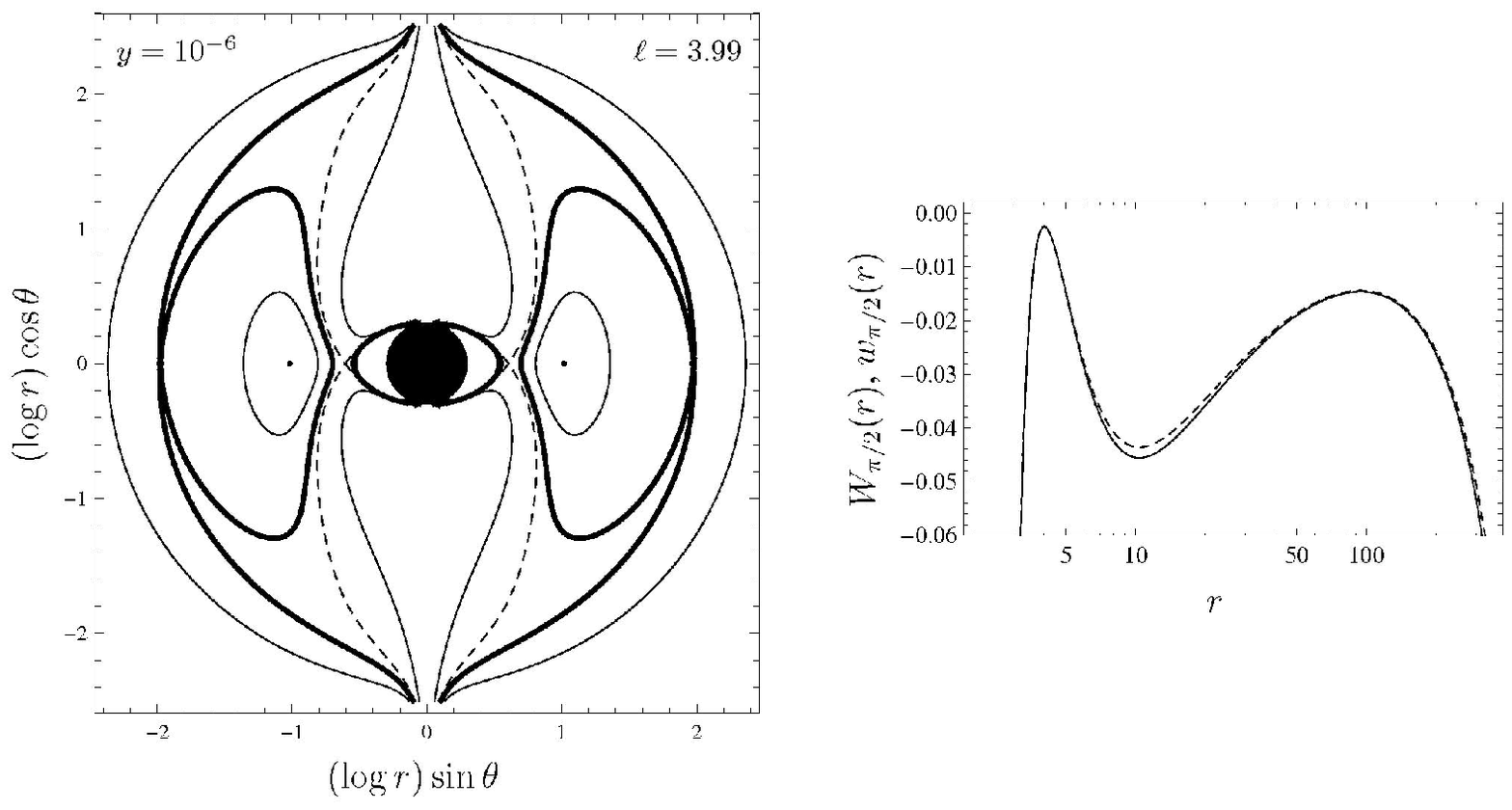}
\par\centering \small(c)\enspace
\end{minipage}
\vskip2ex
\begin{minipage}{1\linewidth}
\centering
\includegraphics[width=.9 \hsize]{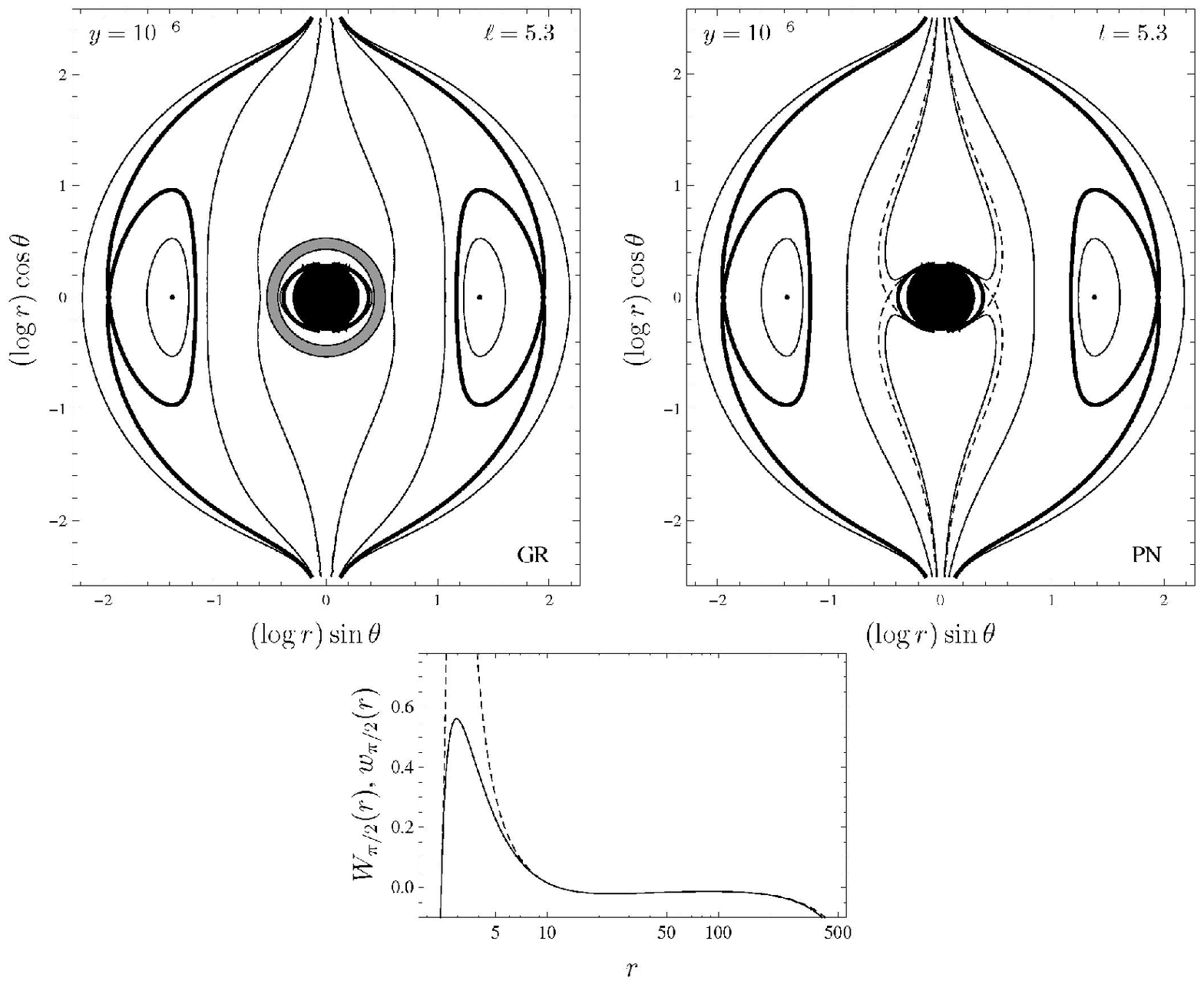}
\par\centering \small(d)\enspace
\end{minipage}
\caption{(Continued.)}
\end{figure}

\subsection{Comparison}
Topology of equipotential surfaces of the potentials $W$ and $w$ can be directly inferred from their behaviour in the equatorial plane, $W_{\pi/2}(r;y)$ and $w_{\pi/2}(r;y)$. Note that the local extrema of both the functions are governed by the same function $\ell_{\rm c}(r;y)$. 
Perfect fluid follows circular geodesics along the radii where the conditions $l=l_{\rm c}(r;y)$ and $\ell=\ell_{\rm c}(r;y)$ are satisfied, which is obvious from the fact that the conditions $\partial_r W_{\rm \pi/2}=0$ and $\partial_r w_{\rm \pi/2}=0$ are equivalent to the conditions of vanishing of pressure gradients.   

\begin{figure}
\begin{minipage}{.49 \linewidth}
\centering
\includegraphics[width=.99 \hsize]{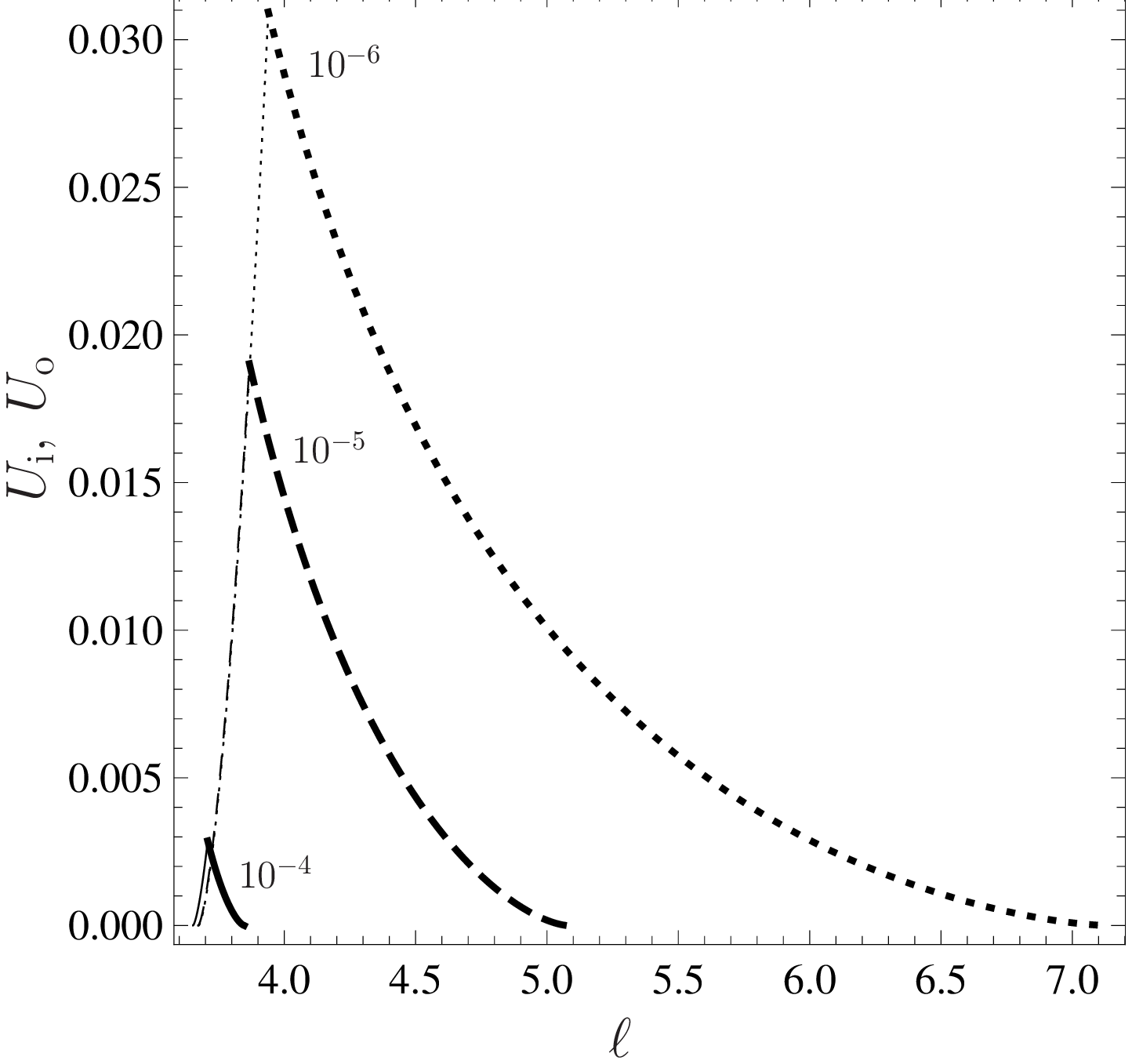}
\par\centering \small(a)
\end{minipage}\hfill
\begin{minipage}{.49 \linewidth}
\centering
\includegraphics[width=.99 \hsize]{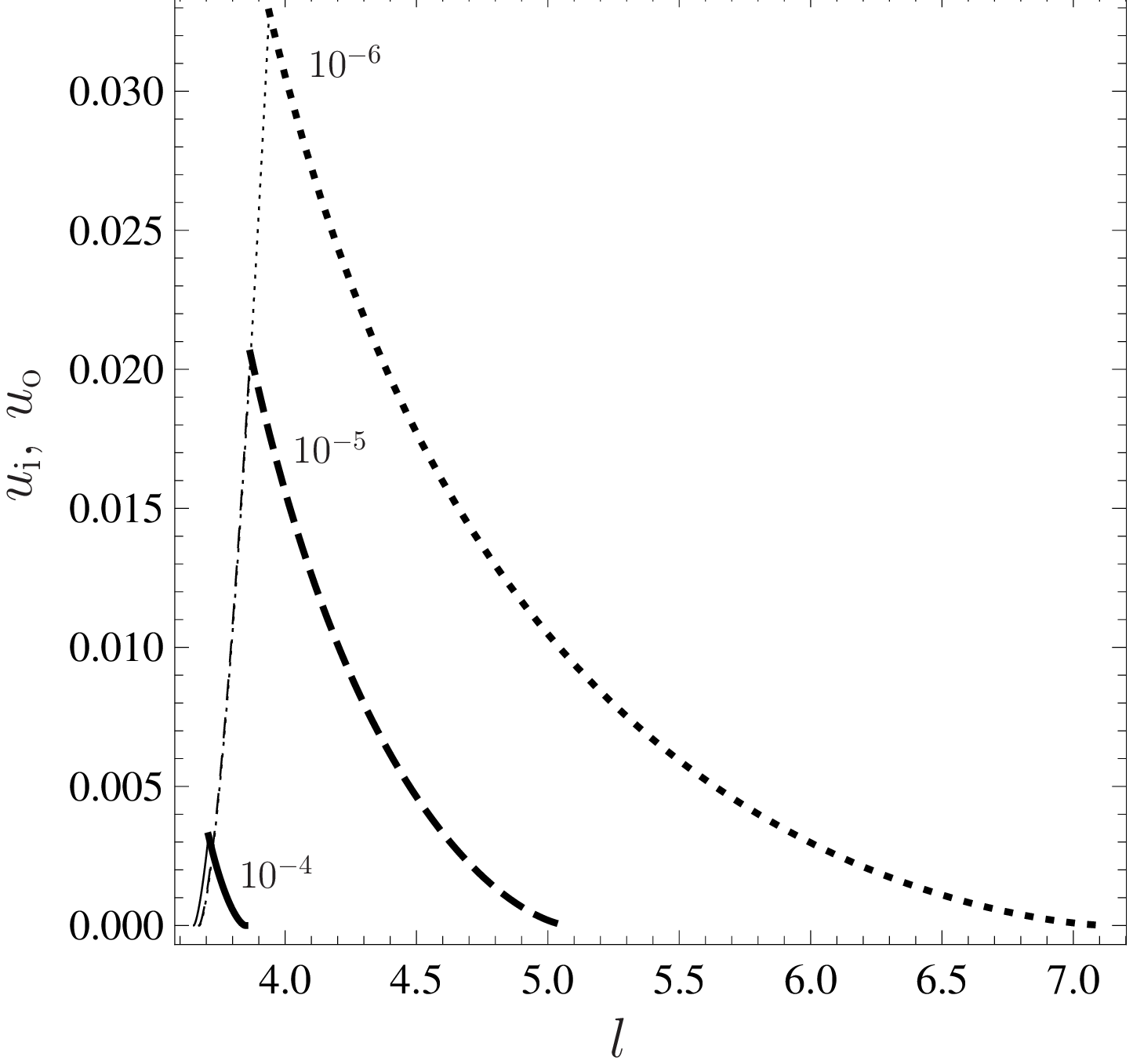}
\par\centering \small(b)
\end{minipage}
\caption{Potential barriers of (a) relativistic thick discs in dependence on the specific angular momentum $\ell\in(\ell_{\rm ms(i)},\ell_{\rm ms(o)})$ and (b) pseudo-Newtonian thick discs in dependence on the angular momentum per unit mass $l\in(l_{\rm ms(i)},l_{\rm ms(o)})$ for three fixed values of the cosmological parameter $y$: $10^{-6}$ (dotted), $10^{-5}$ (dashed) and $10^{-4}$ (solid). Accretion/excretion discs are represented by thin/thick curves, the maxima correspond to marginally bound accretion discs with $\ell=l=\ell_{\rm mb}$.}
\label{f5}
\end{figure}

Our construction of the PN gravitational potential $\psi$ guarantees that the function $\ell_{\rm c}^2(r;y)$ describes the specific angular momentum of freely moving  particles in both the GR and PN approach. Therefore, we can expect close similarities of the character of equipotential surfaces constructed in both approaches. It can be even shown that the shapes of the equipotential surfaces constructed in GR and PN approaches coincide (figure~\ref{f4}). Really, the equipotential surfaces $W(r,\theta)={\rm const}$ and $w(r,\theta)={\rm const}$ are determined by the same relation 
\begin{eqnarray}                                                              \label{e31}
\frac{\d\theta}{\d r} &=&
-\frac{\p_r W}{\p_{\theta}W}=-\frac{\p_r w}{\p_{\theta}w} \\
&=& \frac{r^3(1-yr^3)\sin^2\theta-(r-2-yr^3)^2 \ell^2}{r(r-2-yr^3)^2 \ell^2}\tan\theta, \nonumber
\end{eqnarray}
following  from the conditions $\d W=0$ and $\d w=0$. Using the Schwarzschild coordinates in GR approach and spherical coordinates in PN approach, the equipotential surfaces have the same shape determined by the relation
\begin{equation}                                                              \label{e32}
\sin^2\theta=\left[\frac{r^3}{\ell^2(r-2-yr^3)}+r^2C\right]^{-1}, 
\end{equation}
where the integration constant can be expressed as 
\begin{equation}                                                              \label{e33}
C=\frac{1}{R_{\rm in}^2}-\frac{R_{\rm in}}{\ell^2(R_{\rm in}-2-yR_{\rm in}^3)}.
\end{equation} 
In the PN approach we assume a flat space, while in the GR approach, we have to take the spacetime curvature into acount. Therefore, the proper space distances inside the PN and GR tori differ and we can expect distinct potential barriers (potential differences in the center and in the edge of the torus) or different total masses of the torus, as predicted by the GR and PN approaches for otherwise identical conditions. 

\begin{figure}
\begin{minipage}{.49 \linewidth}
\centering
\includegraphics[width=.99 \hsize]{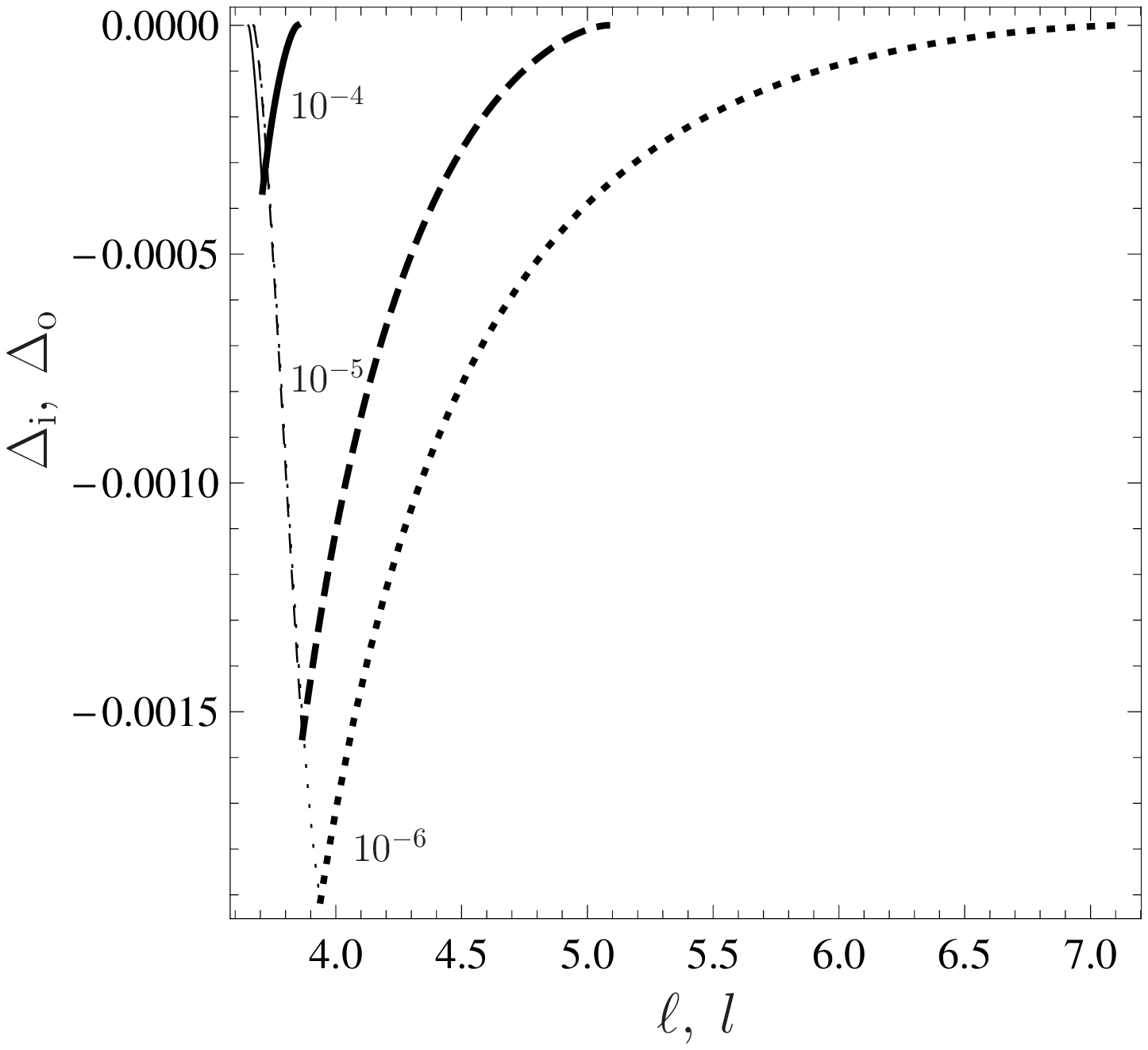}
\par\centering \small(a)
\end{minipage}\hfill
\begin{minipage}{.49 \linewidth}
\centering
\includegraphics[width=.99 \hsize]{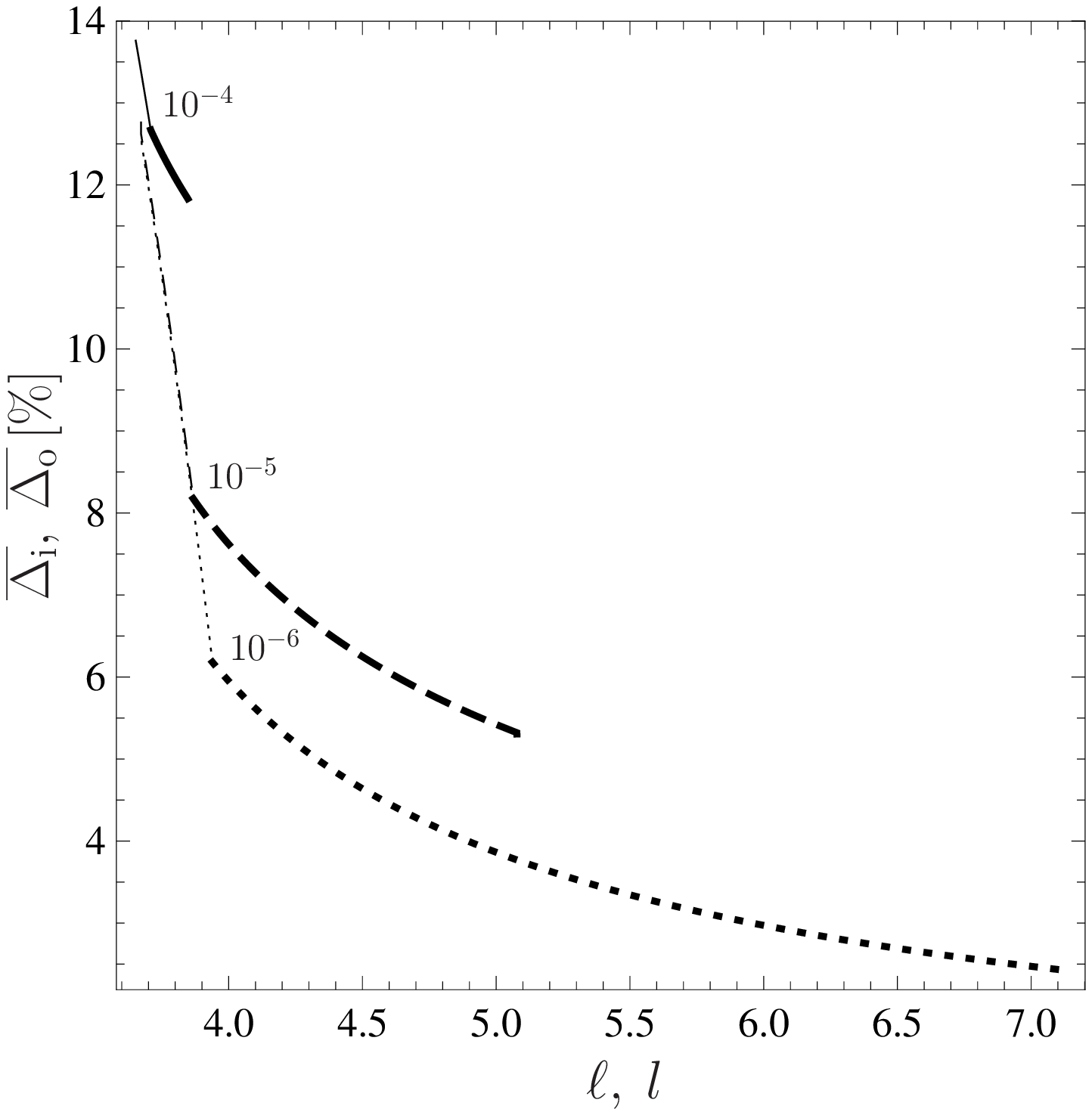}
\par\centering \small(b)
\end{minipage}
\caption{(a) absolute and (b) relative differences of GR and PN potential barriers in accretion (thin) and excretion (thick) discs for three fixed values of the cosmological parameter $y$: $10^{-6}$ (dotted), $10^{-5}$ (dashed) and $10^{-4}$ (solid).}
\label{f6}
\end{figure}

The potential depths comparison can be realized in the following way: determining values of the potentials $W$ and $w$ at their extrema in the equatorial plane, i.e., their inner maxima, $W_{\rm ext,i}$ and $w_{\rm ext,i}$, located at the inner cusp of equipotential surfaces, minima, $W_{\rm ext,c}$ and $w_{\rm ext,c}$, located at the central ring of equipotential surfaces, and outer maxima, $W_{\rm ext,o}$ and $w_{\rm ext,o}$, located at the outer cusp of equipotential surfaces, we can determine the related GR and PN inner and outer potential barriers (figure~\ref{f5})
\bea
U_{\rm i} &=& \left[W_{\rm ext,i}(\ell;y)-W_{\rm ext,c}(\ell;y)\right],         \label{e34a} \\
u_{\rm i} &=& \left[w_{\rm ext,i}(l;y)-w_{\rm ext,c}(l;y)\right],               \label{e34b}
\eea
which represent accretion discs,
\bea
U_{\rm o} &=& \left[W_{\rm ext,o}(\ell;y)-W_{\rm ext,c}(\ell;y)\right],         \label{e35a} \\
u_{\rm o} &=& \left[w_{\rm ext,o}(l;y)-w_{\rm ext,c}(l;y)\right],               \label{e35b}
\eea
which represent excretion discs, and their differences
\begin{eqnarray}
\Delta_{\rm i} &=& U_{\rm i}(\ell;y)-u_{\rm i}(l;y),                            \label{e36a} \\
\Delta_{\rm o} &=& U_{\rm o}(\ell;y)-u_{\rm o}(l;y),                            \label{e36b}
\end{eqnarray}
characterizing variances in the GR and PN descriptions of the accretion and excretion discs. Although the maximal absolute differences $|\Delta_{\rm i,o}|$ grow with the cosmological parameter descending, the relative (percentual) differences $\overline{\Delta}_{\rm i,o}=|\Delta_{\rm i,o}/U_{\rm i,o}|\times 100\%$ behave in opposite way (figure~\ref{f6}).

\section{Adiabatic perfect fluid tori}
\label{s4}
Important physical information is reflected by the behaviour of the fluid inside the toroidal configurations. In realistic accretion discs, it is a~very complex problem how to treat all relevant phenomena concerning properties of the fluid and generation of heat and radiation. Here, in order to compare the GR and PN approaches, we restrict our attention to a~very simple case of adiabatic discs (with negligible influence of radiation).

We assume a perfect fluid obeying an adiabatic pressure--mass-density relation
\begin{equation}                                                           \label{e46}
p=K\rho^{1+1/n},
\end{equation}
where $K$ and $n$ are the adiabatic constant and adiabatic index, respectively. Further, we assume that the pressure is given by the relation for an ideal gas
\begin{equation}                                                           \label{e47}
p=\frac{\rho}{\mu m_{\rm u}}k_{\rm B}T
\end{equation} 
where $T$ is the thermodynamic temperature, and constants $\mu,\,m_{\rm u}$ and $k_{\rm B}$ are mean molecular weight, atomic mass unit and the Boltzmann constant, respectively. In this case (no radiation), the energy density is given by the relation
\be                                                                        \label{e47a}
\epsilon=\rho+np.
\ee

\subsection{General relativistic tori}
Solving equation (\ref{e18}) for the adiabatic fluid, we can express distribution of all basic thermodynamic quantities throughout the toroidal configuration in terms of the potential $W$ in the following way:
\begin{eqnarray}
	\rho &=& \left [\frac{\e^{\Delta W}-1}{K(n+1)}\right ]^n, \quad \Delta W=W_{\rm in}-W,                                                                                        \label{e48} \\
	p &=& K\rho^{1+1/n}=K^{-n}\left (\frac{\e^{\Delta W}-1}{n+1}\right )^{n+1},	  \label{e49} \\
	\epsilon &=& \rho\left[1+\frac{n}{n+1}\left({\e}^{\Delta W} -1\right)\right]  \label{e49a} \\
	&=& \left[\frac{\e^{\Delta W}-1}{K(n+1)}\right ]^n \left[1+\frac{n}{n+1}\left({\e}^{\Delta W}-1 \right)\right]                                                             \label{e49b} \\
  T &=& \frac{\mu m_{\rm u}}{k_{\rm B}}K\rho^{1/n}=\frac{\mu m_{\rm u}}{k_{\rm B}}\frac{\e^{\Delta W}-1}{n+1}.                                                 \label{e50}
\end{eqnarray}
In adiabatic tori, the equipotential surfaces $W(r,\,\theta)=\mbox{const}$ also determine (and are identical) the surfaces of constant thermodynamic quantities $\rho,\,\epsilon,\,p$ and $T$. Extremal values of the thermodynamic quantities are reached at the centre of the tori. 

For a given mean molecular weight $\mu$, which depends on the matter of the disc, relation (\ref{e50}) enables to express the temperature in the centre of the torus in the form
\begin{equation}                                                           	\label{e51}
T_{\rm c}=\frac{\mu m_{\rm u}}{k_{\rm B}}\frac{\e^{\Delta W_{\rm c}}-1}{n+1}, \quad 
\Delta W_{\rm c}=W_{\rm in}-W_{\rm c}.
\end{equation}
Temperature profile of the torus is thus given by
\begin{equation}                                                            \label{e52}
T=T_{\rm c}\frac{\e^{\Delta W}-1}{\e^{\Delta W_{\rm c}}-1}.
\end{equation}

Specifying the central mass-density $\rho_{\rm c}$, the adiabatic constant $K$, the central energy density $\epsilon_{\rm c}$ and the central pressure $p_{\rm c}$ are given by the relations
\begin{eqnarray}
K &=& \rho_{\rm c}^{-1/n}\frac{\e^{\Delta W_{\rm c}}-1}{n+1},               \label{e53} \\
\epsilon_{\rm c} &=& \rho_{\rm c}\left[1+\frac{n}{n+1}\left({\e}^{\Delta W_{\rm c}} -1\right)\right],                                                            \label{e53a} \\
p_{\rm c} &=& \rho_{\rm c}\frac{\e^{\Delta W_{\rm c}}-1}{n+1}.              \label{e54}
\end{eqnarray}
In terms of $\rho_{\rm c}$, which is the only free parameter in description of adiabatic tori, the mass-density profile is given by
\begin{equation}                                                            \label{e55}
\rho=\rho_{\rm c}\left( \frac{\e^{\Delta W}-1}{\e^{\Delta W_{\rm c}}-1} \right)^{n}.
\end{equation}
Similarly, the pressure profile is described by the relation
\be                                                                         \label{e55a}
p=p_{\rm c}\left( \frac{\e^{\Delta W}-1}{\e^{\Delta W_{\rm c}}-1} \right)^{n+1}.
\ee

Central values of the thermodynamic quantities are given by the potential difference between the edge and the centre of the torus, $\Delta_{\rm c}=W_{\rm in}-W_{\rm c}$, which grows with the $\ell=\mbox{const}$ growing from $\ell_{\rm ms,i}$ to $\ell_{\rm ms,o}$. For $\ell=\ell_{\rm mb}$, the toroidal structure is the most extended one and the central values of the thermodynamic quantities reach their extreme values in a fixed BH background. Analogical statements as in GR approach hold, of course, in the PN approach.

\subsection{Pseudo-Newtonian tori}
Starting with relation (\ref{e27}), we can express distribution of all basic thermodynamic quantities throughout the adiabatic torus in terms of the PN potential $w$ in the following way:
\bea
  \rho=\left [\frac{w_{\rm in}-w}{K(n+1)}\right ]^n,                        \label{e61} \\
	p=K^{-n}\left (\frac{w_{\rm in}-w}{n+1}\right )^{n+1},	                  \label{e62} \\
	T=\frac{\mu m_{\rm u}}{k_{\rm B}}\frac{w_{\rm in}-w}{n+1}.                \label{e63}
\eea  
Again, the equipotential surfaces $w(r,\theta)=\mbox{const}$ are identical with surfaces of constant thermodynamic quantities that reach maximal values at the centre of the torus.

The temperature profile of the torus is described by the relation
\be                                                                         \label{e64}
T=T_{\rm c}\frac{w_{\rm in}-w}{w_{\rm in}-w_{\rm c}},
\ee
where the central temperature of the torus is given by
\be                                                                         \label{e65}
T_{\rm c}=\frac{\mu m_{\rm u}}{k_{\rm B}}\frac{w_{\rm in}-w_{\rm c}}{n+1}.
\ee

In terms of the mass-density in the centre $\rho_{\rm c}$, the adiabatic constant $K$, the central pressure $p_{\rm c}$ and the mass-density profile are given by the relations
\bea
K &=& \rho_{\rm c}^{-1/n}\frac{w_{\rm in}-w_{\rm c}}{n+1},                      \label{e66}\\
p_{\rm c} &=& \rho_{\rm c}\frac{w_{\rm in}-w_{\rm c}}{n+1},                     \label{e67}\\
\rho &=& \rho_{\rm c}\left( \frac{w_{\rm in}-w}{w_{\rm in}-w_{\rm c}} \right)^{n}.  \label{e68}
\eea
The pressure profile is then given by
\be
p=p_{\rm c}\left( \frac{w_{\rm in}-w}{w_{\rm in}-w_{\rm c}} \right)^{n+1}.  \label{e68a}
\ee

\subsection{Comparison}

\begin{figure}
\begin{minipage}{1\linewidth}
\includegraphics[width=.78 \hsize]{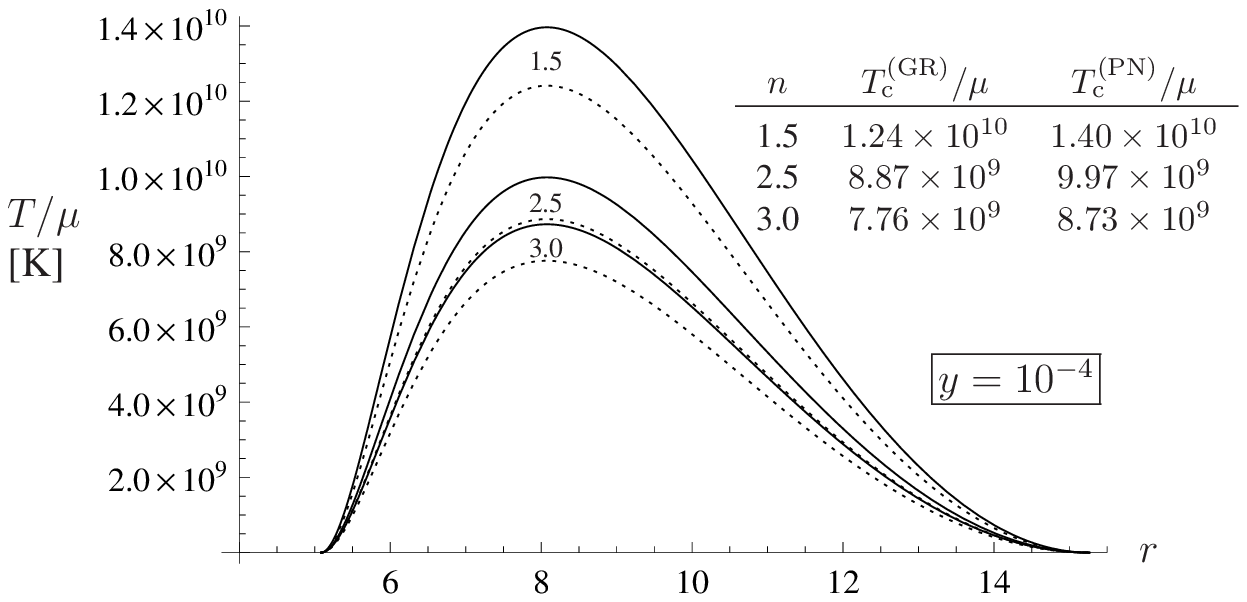}
\end{minipage}
\vskip3ex
\begin{minipage}{1\linewidth}
\includegraphics[width=.75 \hsize]{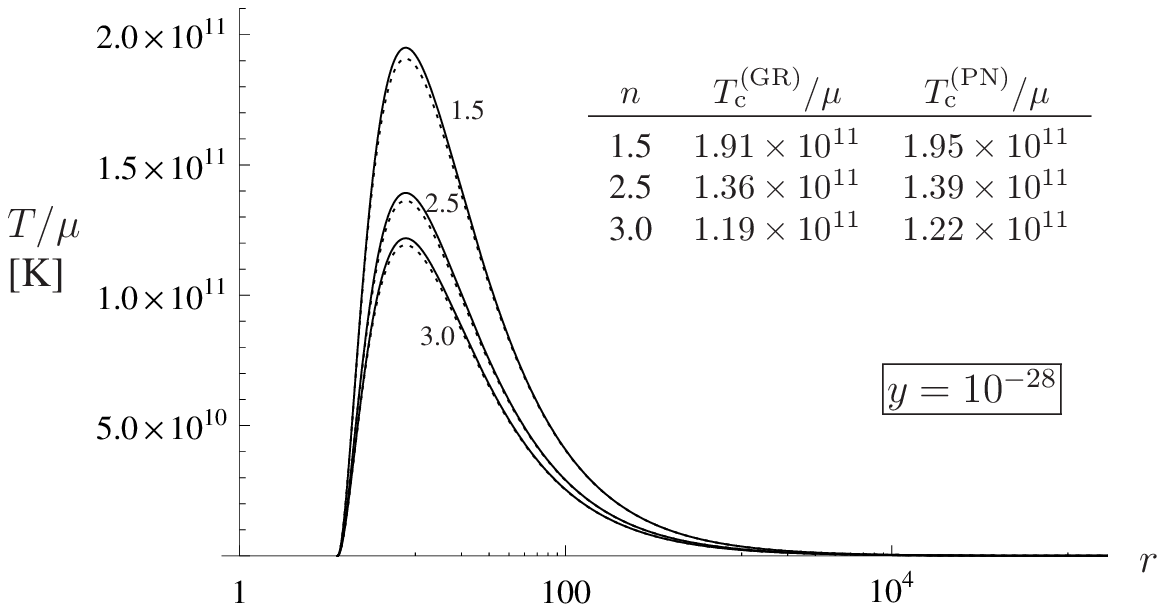}
\end{minipage}
\caption{PN (solid) and GR (dashed) temperature profiles of adiabatic marginally bound accretion discs in the equatorial plane. The tables give a GR and PN temperature in the centre of the disc for a given adiabatic index $n$.}
\label{f7}
\end{figure}

\begin{figure}
\begin{minipage}{1\linewidth}
\includegraphics[width=.95 \hsize]{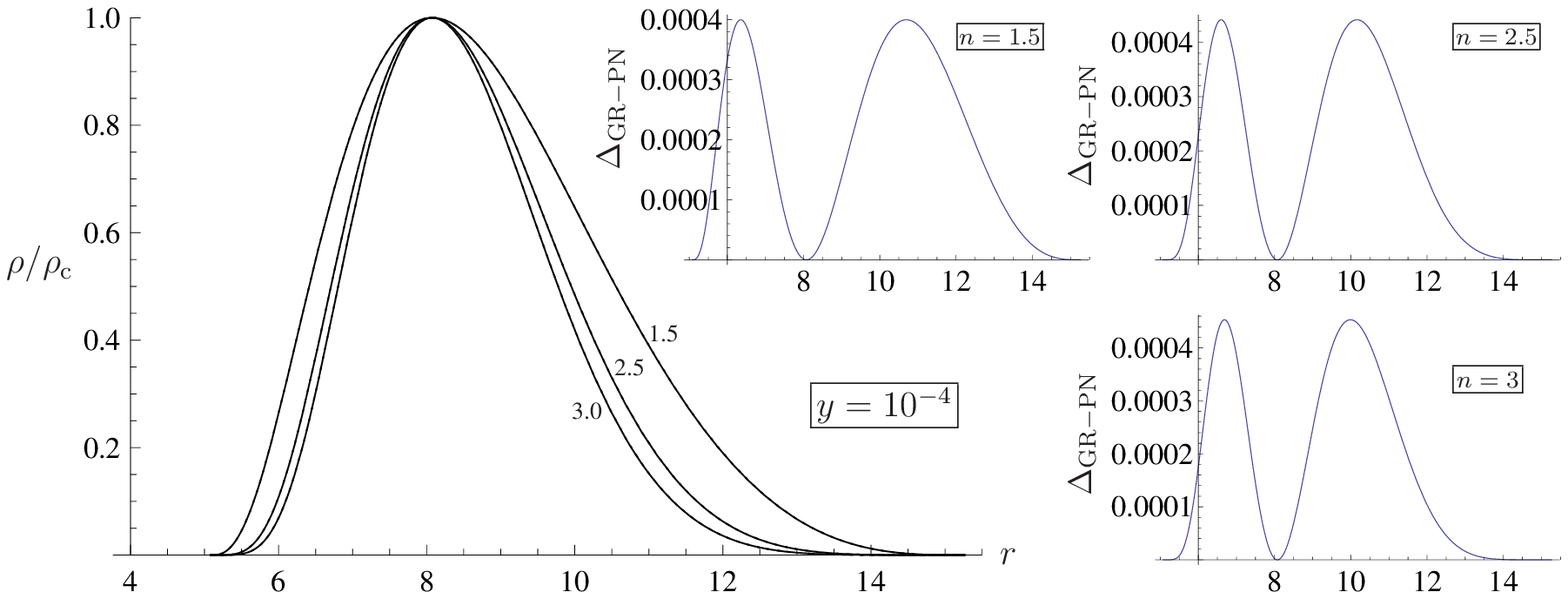}
\end{minipage}
\vskip3ex
\begin{minipage}{1\linewidth}
\includegraphics[width=.95 \hsize]{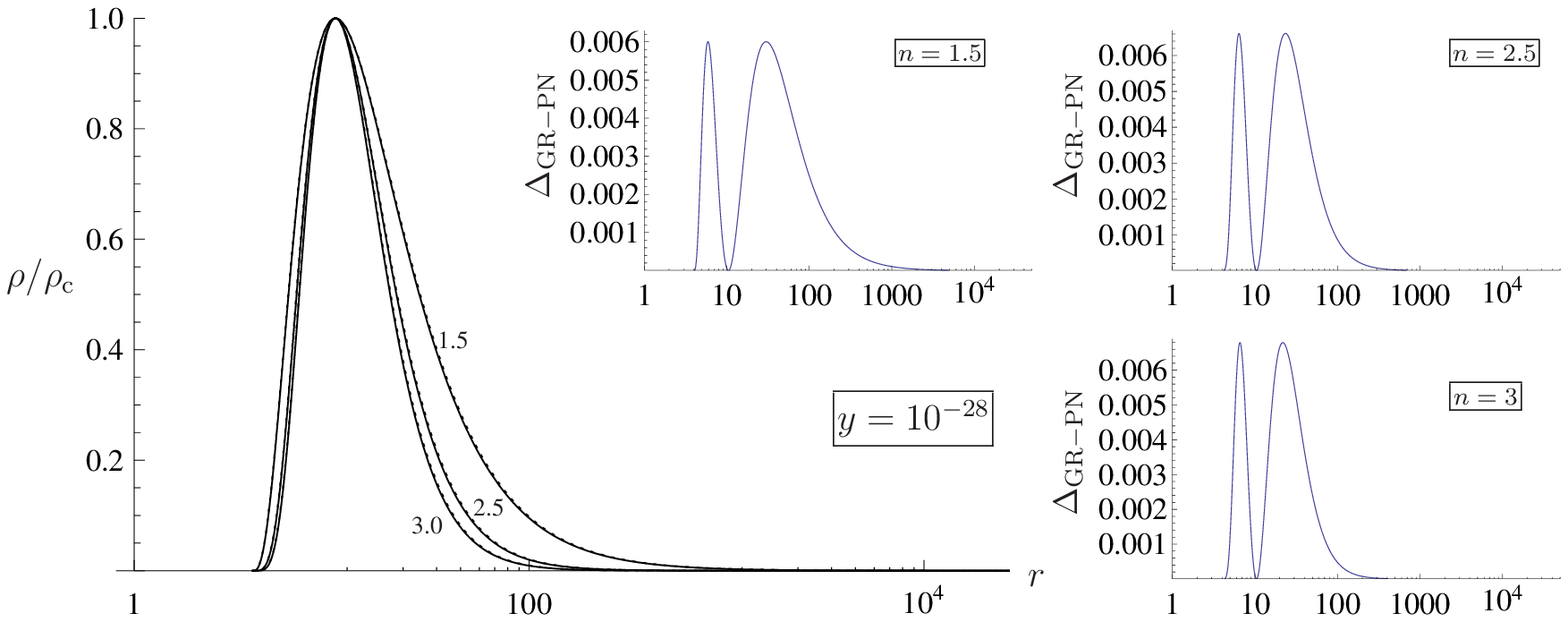}
\end{minipage}
\caption{Equatorial PN and GR mass-density profiles of adiabatic marginally bound accretion discs in units of the central density $\rho_{\rm c}$. Difference of the GR and PN values of $\rho/\rho_{\rm c}$ (denoted as $\Delta_{\rm GR-PN}$) is shown by small figures for various values of the adiabatic index $n$.}
\label{f8}
\end{figure}

\begin{figure}
\begin{minipage}{1\linewidth}
\centering
\includegraphics[width=.72 \hsize]{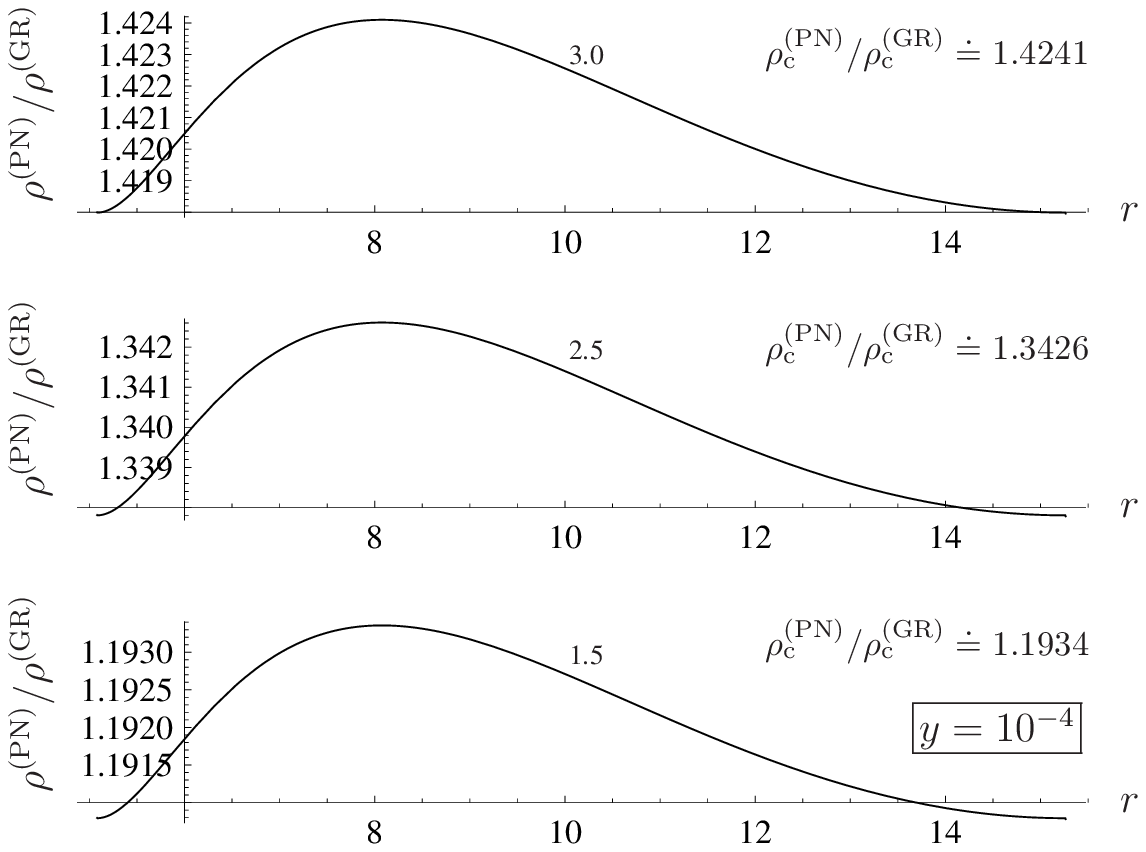}
\end{minipage}
\vskip3ex
\begin{minipage}{1\linewidth}
\centering
\includegraphics[width=.72 \hsize]{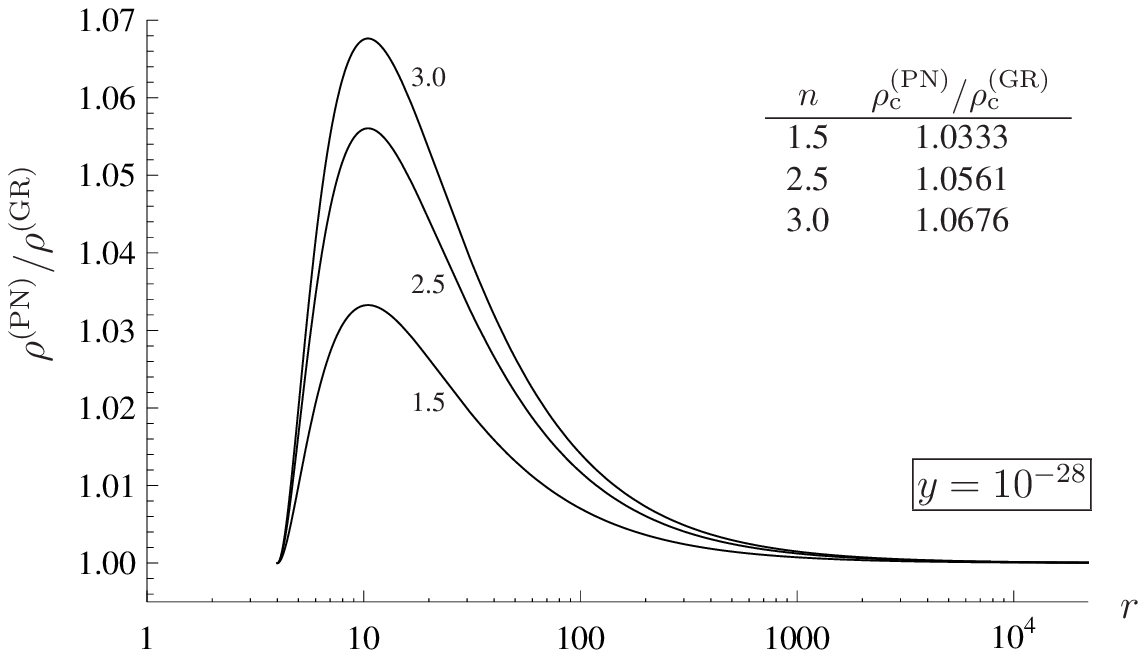}
\end{minipage}
\caption{PN to GR mass-density ratios. Central values for a given adiabatic index $n$ are evaluated.}
\label{f9}
\end{figure}

In order to compare the PN and GR descriptions of adiabatic tori, we plot the temperature and mass-density profiles of the GR and PN marginally bound accretion discs (tori with $\ell(r,\theta)=\ell_{\rm mb}$) in the equatorial plane ($\theta=\pi/2$) for various values of the adiabatic index $n$ and cosmological parameter $y$ (figures \ref{f7} and \ref{f8}). Maximal values of $T$ and $\rho$ correspond to the centre of the torus. We can see that the PN-temperature profile is little bit higher than the GR-temperature profile and the difference of these values at a given $r$ grows with growing cosmological parameter $y$. On the other hand, the PN density profile $\rho/\rho_{\rm c}$ is only slightly lower than the GR density profile, and the difference of these values descends with the cosmological parameter growing. The difference is demonstrated by the quantity $\Delta_{\rm GR-PN}$ in figure \ref{f8}. It should be noted, however, that when the adiabatic constant $K$ is the same for both PN and GR discs, the mass-density predicted by the PN approach (relation (\ref{e61})) is always higher than the mass-density given by the GR approach (relation (\ref{e48})) at a given $r$. Their difference is the biggest in the centre of the torus. Ratio of the PN and GR predictions of mass densities in adiabatic tori is shown in figure \ref{f9}. 

Similarly, we can also plot pressure profiles of GR and PN adiabatic discs for various values of the adiabatic index $n$ and cosmological parameter $y$, described by relations (\ref{e55a}) and (\ref{e68a}). For given values of $y$ and $n$, the profiles $p/p_{\rm c}$ show the same behaviour as mass-density profiles for the same cosmological parameter $y$ but adiabatic index $(n-1)$. Thus, the pressure profiles are narrower than corresponding mass-density profiles.

\begin{figure}
\begin{minipage}{1\linewidth}
\centering
\includegraphics[width=.75 \hsize]{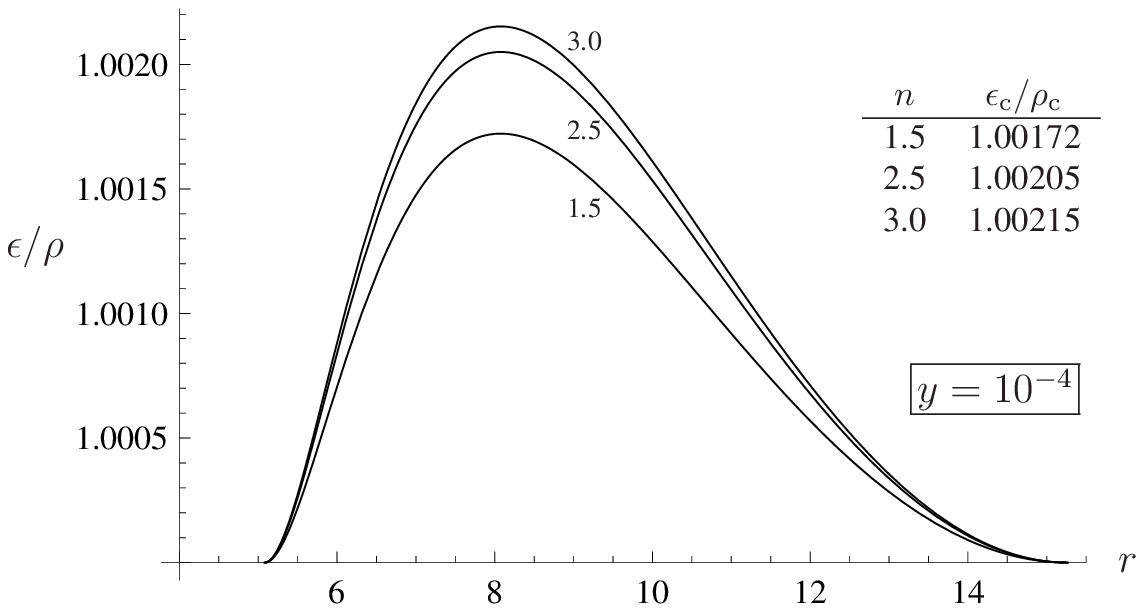}
\end{minipage}
\vskip3ex
\begin{minipage}{1\linewidth}
\centering
\includegraphics[width=.75 \hsize]{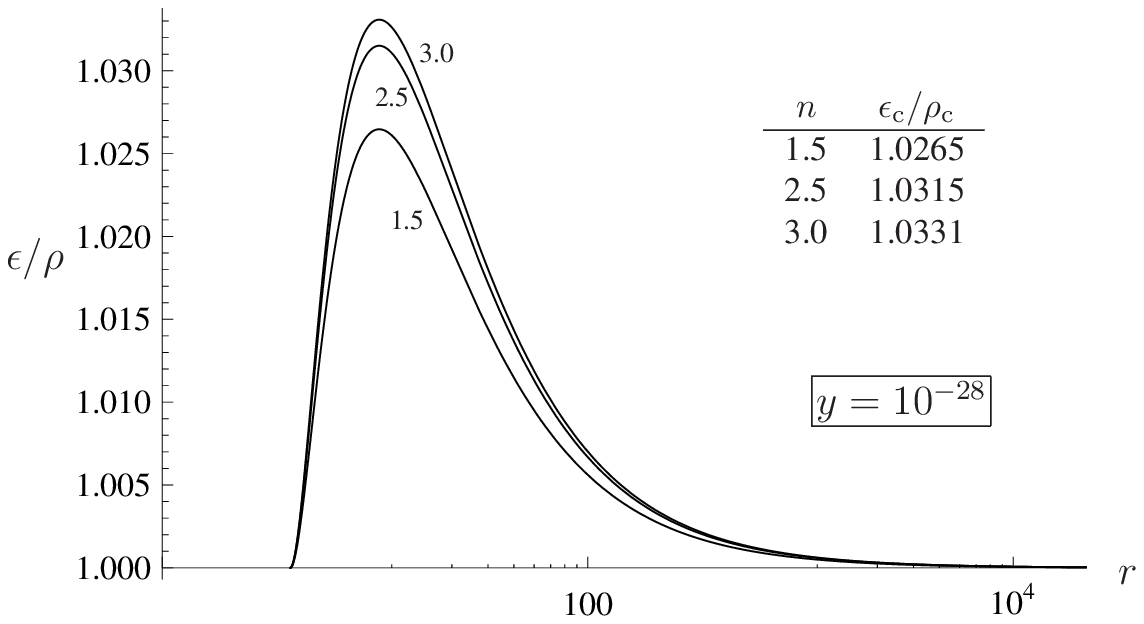}
\end{minipage}
\caption{Energy density to mass-density ratios (in units $c=1$). The tables give values in the centre of the disc for a given adiabatic index $n$.}
\label{f10}
\end{figure}

In the framework of general relativity, it is possible to analyze how much relativistic is the fluid matter in the adiabatic torus in dependence on the adiabatic index $n$ and cosmological parameter $y$, investigating relation (\ref{e49a}) for the special energy density. The results are shown in figure \ref{f10}. We can see that the importance of the GR effects in the torus grows with the adiabatic index $n$ growing and the cosmological parameter $y$ descending. However, the special- relativity effect is important only in the inner part of the disc around its centre, and the total energy density in the centre, where it is maximal, exceeds the rest-energy density maximally by $\sim 3\%$ of the rest value.

\section{Total mass of adiabatic tori}
\label{s5}
In this section, we compare GR and PN results giving a total mass of the adiabatic perfect fluid torus. Using this quantity, we can put some limit on validity of the test-disc approximation implicitly assumed in our calculations. 

\subsection{General relativistic mass formula}
Total mass of a~torus is determined by the Tolman formula \cite{Koz-Jar-Abr:1978:ASTRA:}
\begin{equation}                                                            \label{e56}
m=\int_{\rm disc}(-T^t_t+T^r_r+T^{\theta}_{\theta}+T^{\phi}_{\phi})\sqrt{-g}\,\d r\d\theta\d\phi,
\end{equation}
where $T^i_k$ are components of the perfect fluid stress-energy tensor and $g=\mbox{det}(g_{ik})$. For a perfect fluid torus orbiting a SdS black hole, the Tolman formula reads
\begin{equation}                                                            \label{e57}
m=\int_{\rm disc}\left[ \frac{1+\Omega\ell}{1-\Omega\ell}(\epsilon+p)+2p\right]r^2\sin\theta\,\d r\d\theta\d\phi,	
\end{equation}
where the angular velocity profile $\Omega(r,\,\theta)$ is determined by the spacetime geometry and the specific angular momentum distribution $\ell(r,\,\theta)$ through the relation (\ref{e17}):  
\be                                                                         \label{e58}
\Omega(r,\,\theta)=\frac{r-2-yr^3}{r^3\sin^2\theta}\ell(r,\,\theta).
\ee

In the case of the adiabatic tori characterized by the adiabatic index $n$ and central density $\rho_{\rm c}$, relation (\ref{e57}) can be rewritten to the form
\bea                                                                        \label{e59}
m &=& \rho_{\rm c}\int_{\rm disc}\left[ \left( \frac{1+\Omega\ell}{1-\Omega\ell}+\frac{2}{n+1}\right){\e}^{\Delta W}-\frac{2}{n+1}\right] \nonumber \\ 
& & \times\left( \frac{\e^{\Delta W}-1}{\e^{\Delta W_{\rm c}}-1} \right)^{n}r^2\sin\theta\,\d r\d\theta\d\phi,
\eea
enabling to find the mass of the adiabatic torus from the behaviour of the potential field $W(r,\,\theta)$ in the disc. In the limit of non-relativistic fluid for which $p\ll\epsilon \approx\rho$, the mass of the adiabatic torus is approximatively given by
\be                                                                         \label{e60}
m\approx\rho_{\rm c}\int_{\rm disc}\frac{1+\Omega\ell}{1-\Omega\ell}\left( \frac{W_{\rm in}-W}{W_{\rm in}-W_{\rm c}} \right)^{n}r^2\sin\theta\,\d r\d\theta\d\phi.
\ee

\subsection{Pseudo-Newtonian mass formula}
In the Newtonian description of the fluid, we naturally assume a~non-relativistic perfect fluid with $p\ll\rho$. Geometry of the spacetime is Euclidean.

The mass of a~torus can be determined from the relation
\be                                                                         \label{e69}
m=\int_{\rm disc}\rho(r,\theta,\phi)r^2\sin\theta\,\d r\d\theta\d\phi.
\ee 
Using the results of the previous section for adiabatic PN tori, the mass of such a~torus is given by the formula
\be                                                                         \label{e70}
m=\rho_{\rm c}\int_{\rm disc}\left(\frac{w_{\rm in}-w}{w_{\rm in}-w_{\rm c}}\right)^n r^2\sin\theta\,\d r\d\theta\d\phi.
\ee 

\subsection{Comparison}
Information on mass of the torus provides an opportunity to discuss the relevance of the test-disc approximation. It is not trivial to find a reliable criterion in this case because of large extension of the torus in comparison with the BH extension. A very simple criterion is to compare the total mass of the torus with mass parameter of the black hole, stating that the test-disc approximation is acceptable for $m\ll M$. Due to much smaller `compactness' of the torus, as compared with the black hole, we assume $m\leq M/10$ to be sufficient enough for the test-disc approximation, taking $m=M/10$ as a testing limit. Central mass densities of adiabatic tori with $m=M/10$ are presented in table~\ref{t2}. Thus the test-disc approximation should be relevant for adiabatic tori with central mass densities lower than those in the tables (for given adiabatic index $n$ and mass of the black hole $M$). 

In order to compare the GR and PN masses of adiabatic tori, we calculate in both the approaches the central mass densities, corresponding to the same mass of the torus. At first, we choose the marginally bound accretion discs (which correspond to the most extended toroidal structures in any given spacetime), i.e., the discs with $\ell=\ell_{\rm mb}$, and calculate the mass densities in the centre for the limiting (test-disc) situation when  the total mass of the torus is given by  $10m=M=\sqrt{3y/\Lambda}=1$ in dimensionless formulation. 

\begin{figure}
\centering
\includegraphics[width=0.75\hsize]{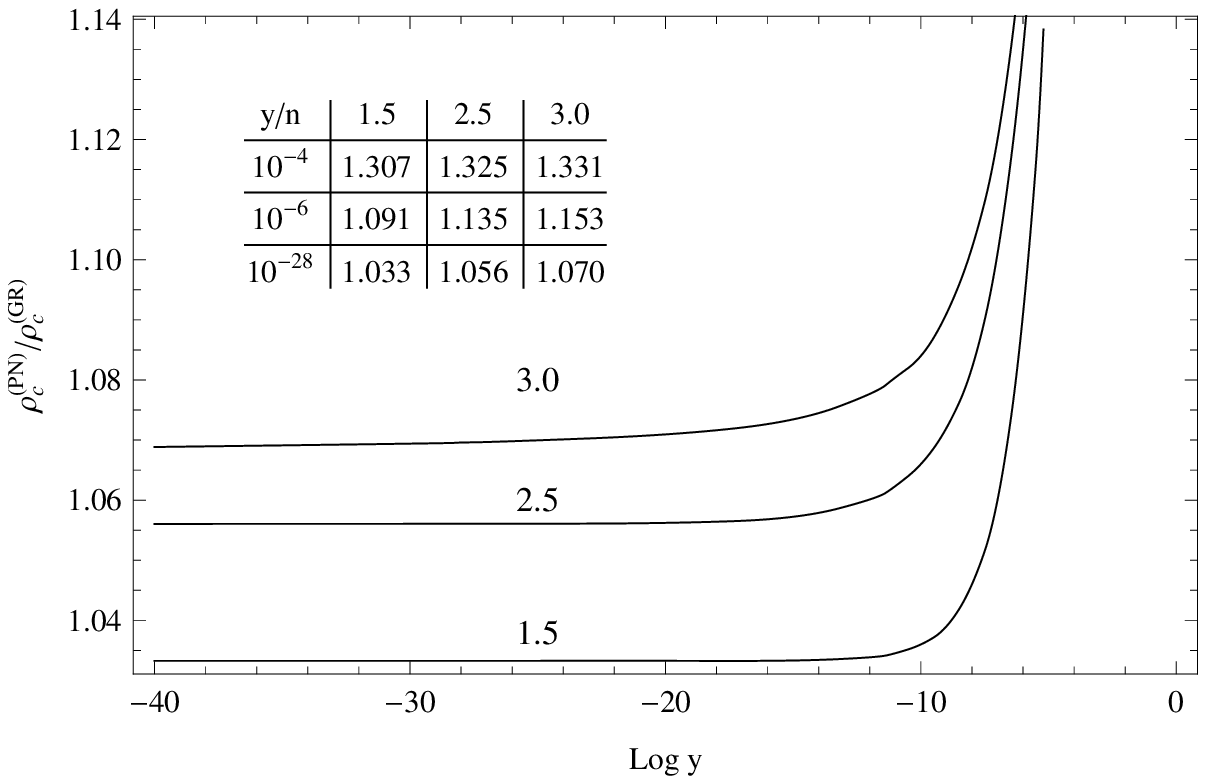}
\caption{Ratio of central mass densities of the PN and GR adiabatic marginally bound accretion discs of the same mass equal to the mass of the black hole, i.e. $m=M$, for three values of the adiabatic index $n=(1.5,2.5,3.0)$ as a function of the cosmological parameter $y$. Attached table shows the ratios for three representative values of $y$.}
\label{f11}
\end{figure}

With respect to the current value of the cosmological constant ($\Lambda_0 \doteq 1.3\times 10^{-56}$\,cm$^{-2}$) and recent observations confirming existence of black holes with masses up to $M\sim 10^{10}$\,M$_{\odot}$ \cite{Zio:2008:CHIJAA:}, the highest current astrophysically relevant values of the cosmological parameter are about $y\sim 10^{-25}$. 
For marginally bound accretion tori orbiting intermediate-mass and supermassive black holes, the above-mentioned central mass-densities (and pressures) are presented in table~\ref{t2} in physical units. Ratio of such PN and GR central mass-densities in dependence on the cosmological parameter $y$ is presented in figure~\ref{f11} for three typical values of the adiabatic index $n$. Notice that for astrophysically relevant black holes with $y<10^{-25}$ the precision of the PN approach is relatively high and differences of GR and PN results are few percent only.

Analogical calculations were performed for accretion discs with $\ell=1/2(\ell_{\rm ms,i}+\ell_{\rm mb})$ and excretion discs with $\ell=1/2(\ell_{\rm mb}+\ell_{\rm ms,o})$. The results are again presented in table~\ref{t2}.  

\begin{sidewaystable}
\caption{GR and PN central mass densities and pressures of adiabatic thick discs with total mass $m=M/10$; $M$ is the mass parameter of a~particular SdS black hole. The results are given for current value of the cosmological constant, $\Lambda_0 \doteq 1.3\times 10^{-56}$cm$^{-2}$, and three values of the adiabatic index $n=(1.5,2.5,3.0)$. Size of the disc is characterized by position of the outer cusp of toroidal configuration $R_{\rm out}$ (given in units of M and in parsecs). The inner cusp of accreting configurations is located in vicinity of the black hole at $R_{\rm in}\approx 4{\rm M}$, while the inner edge of excretion discs is located far away from the black hole at $\sim 10^7$M for $y=10^{-26}$, or $\sim 10^{10}$M for $y=10^{-34}$.}
{\small
\begin{tabular}{@{}lllllllllll}
\br
$y$ & $M$ & \centre{2}{$R_{\rm out}$} & $\rho_{\rm c(1.5)}$ & $\rho_{\rm c(2.5)}$ & $\rho_{\rm c(3.0)}$ & $p_{\rm c(1.5)}$ & $p_{\rm c(2.5)}$ & $p_{\rm c(3.0)}$ & \\
& \centre{1}{$[{\rm M_{\odot}}]$} & [M] & [pc] & [${\rm kg/m^{3}}$] & [${\rm kg/m^{3}}$] & [${\rm kg/m^{3}}]$ & [Pa] & [Pa] & [Pa] & \\
\mr
\multicolumn{6}{l}{accretion discs ($\ell=1/2(\ell_{\rm ms,i}+\ell_{\rm mb})\doteq 3.8\,M$)} & & & & & \\
\mr
$10^{-26}$&$10^{10}$&20.6&$9.9\times 10^{-3}$&$1.98\times 
10^{-4}$&$3.32\times 10^{-4}$ & $4.04\times 10^{-4}$&$1.11\times 
10^{11}$&$1.33\times 10^{11}$&$1.42\times 10^{11}$&PN\\
&&&&$1.60\times 10^{-4}$&$2.62\times 10^{-4}$&$3.16\times 10^{-4}$ & 
$8.30\times10^{10}$&$9.71\times 10^{10}$&$1.02\times 10^{11}$&GR\\[1ex]
$10^{-34}$&$10^{6}$&$20.6$&$9.9\times 10^{-7}$&$1.98\times 
10^{4}$&$3.32\times 10^{4}$ & $4.04\times 10^{4}$&$1.11\times 
10^{19}$&$1.33\times 10^{19}$&$1.42\times 10^{19}$&PN\\
&&&&$1.60\times 10^{4}$&$2.62\times 10^{4}$&$3.16\times 10^{4}$& 
$8.30\times10^{18}$&$9.71\times10^{18}$&$1.02\times 10^{19}$&GR\\[1ex]
$10^{-40}$&$10^{3}$&$20.6$&$9.9\times 10^{-10}$&$1.98\times 
10^{10}$&$3.32\times 10^{10}$ & $4.04\times 10^{10}$&$1.11\times 
10^{25}$&$1.33\times 10^{25}$&$1.42\times 10^{25}$&PN\\
&&&&$1.60\times 10^{10}$&$2.62\times 10^{10}$&$3.16\times 10^{10}$& 
$8.30\times10^{24}$&$9.71\times10^{24}$&$1.02\times 10^{25}$&GR\\[1ex]
$10^{-44}$&$10^{1}$&$20.6$&$9.9\times 10^{-12}$&$1.98\times 
10^{14}$&$3.32\times 10^{14}$ & $4.04\times 10^{14}$&$1.11\times 
10^{29}$&$1.33\times 10^{29}$&$1.42\times 10^{29}$&PN\\
&&&&$1.60\times 10^{14}$&$2.62\times 10^{14}$&$3.16\times 10^{14}$& 
$8.30\times10^{28}$&$9.71\times10^{28}$&$1.02\times 10^{29}$&GR\\
\mr
\multicolumn{6}{l}{marginally bound accretion discs ($\ell=\ell_{\rm mb}\doteq 4\,M$)} & & & & & \\
\mr
$10^{-26}$&$10^{10}$&$10^8$&$230\times 10^3$&$3.60\times10^{-16}$&$1.14\times10^{-9}$ & $3.17\times10^{-7}$&$5.84\times 10^{-1}$&$1.32\times 10^{6}$&$3.21\times 10^{8}$&PN\\
&&&&$3.49\times10^{-16}$&$1.08\times10^{-9}$&$2.97\times10^{-7}$ & $5.53\times10^{-1}$&$1.22\times 10^{6}$&$2.94\times 10^{8}$&GR\\[1ex]
$10^{-34}$&$10^{6}$&$10^{11}$&$11\times 10^3$&$3.60\times10^{-12}$&$5.30\times10^{-3}$ & $2.18\times10^1$&$5.84\times10^{3}$&$6.14\times10^{12}$&$2.21\times10^{16}$&PN\\
&&&&$3.49\times10^{-12}$&$5.02\times10^{-3}$&$2.04\times10^1$& $5.53\times10^{3}$&$5.69\times10^{12}$&$2.02\times10^{16}$&GR\\
%
\mr
\multicolumn{9}{l}{excretion discs ($\ell=1/2(\ell_{\rm mb}+\ell_{\rm ms,o})$); $\ell\doteq 7.4\times 10^3\,M$ for $y=10^{-26}$, $\ell\doteq 1.6\times 10^5\,M$ for $y=10^{-34}$} & & \\
\mr
$10^{-26}$&$10^{10}$&$10^{8}$&$210\times 10^3$&$7.077\times 10^{-26}$ & $1.899\times 10^{-25}$&$2.688\times 10^{-25}$&$1.54\times 10^{-17}$&$2.94\times 10^{-17}$ & $3.64\times 10^{-17}$&PN, GR \\[1ex]
$10^{-34}$&$10^{6}$&$10^{11}$&$9.9\times 10^3$&$7.081\times 10^{-26}$ & $1.901\times 10^{-25}$&$2.690\times 10^{-25}$&$3.31\times 10^{-20}$&$6.35\times 10^{-20}$ & $7.86\times 10^{-20}$&PN, GR \\
\br
\end{tabular}
}
\label{t2}
\end{sidewaystable}

We can see that the PN densities and pressures predominate over the GR ones, but both the approaches imply results of the same order. In the case of excretion discs, the results even almost coincide. Apparently, this is because the excretion discs do not take place in `highly-curved' region of the spacetime nearby the black hole. For maximally extended tori ($\ell=\ell_{\rm mb}$), differences of the PN and GR approaches are relatively small (few percent), since only small part of such tori are located in the `highly-curved' region close to the BH horizon. The PN and GR results are of the same order (tens of percent differences) even for accretion tori fully located in the strong gravity region nearby the horizon.
With the cosmological parameter $y$ descending, the precision of PN calculations increases. 

Comparing results in table~\ref{t2}, obtained for the case $m=M/10$, with relations (\ref{e59}) and (\ref{e70}) for mass of the torus in both approaches, we can deduce that in the case of equal central mass densities of PN and GR tori, the mass of PN torus must be a little bit smaller than the mass of its GR counterpart (but of the same order!).\footnote{Since the PN approach is defined in order to describe the BH physics in the Euclidean space, the proper volume of the curved region in the GR approach is greater than its flat PN counterpart.} For example, if we choose marginally bound adiabatic torus with $n=1.5$ and $\rho_{\rm c}=10^{-15}$\,kg/m$^3$, orbiting a supermassive black hole with $M=10^6$\,M$_{\odot}$, the mass of such a torus in the PN approach will be $m_{\rm PN}\doteq 2.8\times 10^{-5}$\,M, while in the GR approach $m_{\rm GR}\doteq2.9\times 10^{-5}$\,M. Again, with decreasing values of the cosmological parameter $y$, precision of the PN calculation increases.

\begin{table}
\caption{GR and PN central temperatures of adiabatic thick discs with total mass $m=M/10$; $M$ is the mass parameter of a~particular SdS black hole. The results are given for current value of the cosmological constant, $\Lambda_0 \doteq 1.3\times 10^{-56}$cm$^{-2}$, and three values of the adiabatic index $n=(1.5,2.5,3.0)$.}
{\small
\begin{tabular}{@{}llllll}
\br
$y$ & $M$ & $T_{\rm c(1.5)}/\mu$ & $T_{\rm c(2.5)}/\mu$ & $T_{\rm c(3.0)}/\mu$ & \\
& [${\rm M_{\odot}}$] & \centre{1}{[K]} & \centre{1}{[K]} & \centre{1}{[K]} & \\
\mr
\multicolumn{3}{l}{accretion discs ($\ell\doteq 3.8\,M$)}& & & \\
\mr
$10^{-40}$--$10^{-26}$&$10^{3}$--$10^{10}$&$6.74\times 10^{10}$&$4.82\times 10^{10}$ &$4.22\times 10^{10}$& PN\\
& & $6.24\times 10^{10}$&$4.46\times 10^{10}$ &$3.90\times 10^{10}$&GR\\
\mr
\multicolumn{4}{l}{marginally bound accretion discs ($\ell=\ell_{\rm mb}\doteq 4\,M$)} & & \\
\mr
$10^{-40}$--$10^{-26}$&$10^{3}$--$10^{10}$&$1.95\times 10^{11}$&$1.39\times 10^{11}$ &$1.22\times 10^{11}$& PN\\
& & $1.91\times 10^{11}$&$1.36\times 10^{11}$ &$1.19\times 10^{11}$& GR\\
\mr
\multicolumn{6}{l}{excretion discs ($\ell\doteq 7.4\times 10^3\,M$ for $y=10^{-26}$, $\ell\doteq 1.6\times 10^5\,M$ for $y=10^{-34}$} \\
\mr
$10^{-26}$&$10^{10}$&$2.61\times 10^4$&$1.86\times 10^4$&$1.63\times 10^4$& PN, GR \\[1ex]
$10^{-34}$&$10^{6}$ & 56.2 & 40.2 & 35.1 & PN, GR \\
\br
\end{tabular}
}
\label{t3}
\end{table}

\section{Discussion}
\label{dis}
We shall discuss behaviour of thermodynamic quantities in three types of toroidal configurations, and present some comments on the definition of the PN potential. 

\subsection{Profiles of thermodynamic quantities}
Central values of mass-density $\rho$, pressure $p$ and temperature $T$ in adiabatic tori, suggested by PN and GR approaches (tables~\ref{t2}, \ref{t3}), are given for current astrophysically relevant values of cosmological parameter $y\in (10^{-44}-10^{-26})$, reflecting the range of BH masses from stellar-mass ($10\,{\rm M_{\odot}}$) to the most supermassive ones ($10^{10}\,{\rm M_{\odot}}$, which is comparable with mass of the heaviest supermassive black hole TON 618 being observed till now) and the present value of the cosmological constant $\Lambda_0\doteq 1.3\times 10^{-56}\,{\rm cm}^{-2}$. The calculations are performed for both accretion and excretion discs, as well as for the most extended marginally bound tori enabling simultaneous outflow of matter into the black hole  and to the outer space. In all  cases we assume the test-disc approximation, with the limiting value $m=M/10$. 

\subsubsection{Most extended tori}
In the marginally bound tori with $\ell=\mbox{const}=\ell_{\rm mb}$ orbiting any astrophysically relevant SdS black hole, the central temperature reaches the extreme value independently of the mass of the torus, since for adiabatic fluid it is fully determined by the potential depths $\Delta W_{\rm c}=W_{\rm in}-W_{\rm c}$ in GR and $\Delta w_{\rm c}=w_{\rm in}-w_{\rm c}$ in PN approaches that reach their maxima for the tori with $\ell=\ell_{\rm mb}$ (cf. equations (\ref{e51}) and (\ref{e65})). Note that for the mentioned interval of $y$ the central temperature is almost independent of $y$ for both GR and PN models, being a few percent higher in the PN model as compared to GR ones. On the other hand, the central values of mass-density and pressure depend on the total mass of the disc (cf. equations (\ref{e59}), (\ref{e70}) and (\ref{e54}), (\ref{e67})).
Extension of $\ell=\ell_{\rm mb}$ tori expressed in terms of gravitational radius of the central black hole $M$ falls with the cosmological parameter $y$ growing. Since for $y\lesssim 10^{-10}$, there is $r_{\rm mb,o}\sim r_{\rm s}=y^{-1/3}M$ and $r_{\rm h}\sim M$, an approximate relation for $r_{\rm mb,o}$ can be given in a simple form $r_{\rm mb,o}/r_{\rm h}\sim M^{-2/3}\Lambda^{-1/3}$. The central mass-density and pressure grow sharply with descending $y$ (BH mass). The PN model predicts the central density and pressure by $\lesssim 10\%$ (and typically only by few percent) higher than the GR model.

\begin{table}
\caption{Position of the outer edge $R_{\rm out}$ of accretion tori in units of $M$ in dependence on the value of constant specific angular momentum (the results are the same for both PN and GR marginally stable tori) and the GR values (of the same order as PN values) of central densities in $\rm kg/m^3$ for $n=1.5$ in the disc which is characterized by parameter $k$ in the relation $\ell=\ell_{\rm ms,i}+k(\ell_{\rm mb}-\ell_{\rm ms,i})$ and value of cosmological parameter $y=10^{-26}$. Note that there is no significant dependence of the inner edge position on the parameter $k$, being located at $R_{\rm in}\doteq 4\,M$. }
\begin{indented}
\item[]\begin{tabular}{@{}llllllll}
\br
$k$ &$0.1$&$0.5$&$0.9$&$0.99$&$0.9999$&$0.999999$&$1$\\
\mr
$R_{\rm out} {\rm [M]}$&$10^0$&$10^1$&$10^2$&$10^3$&$10^5$&$10^7$&$10^8$\\
$\rho_{\rm c} {\rm [kg/m^3]}$&$10^{-2}$&$10^{-4}$&$10^{-6}$&$10^{-7}$&$10^{-10}$&$10^{-13}$&$10^{-16}$\\
\br
\end{tabular}
\end{indented}
\label{t4}
\end{table}

\subsubsection{Accretion tori}
\begin{figure}
\centering
\includegraphics[width=0.75\hsize]{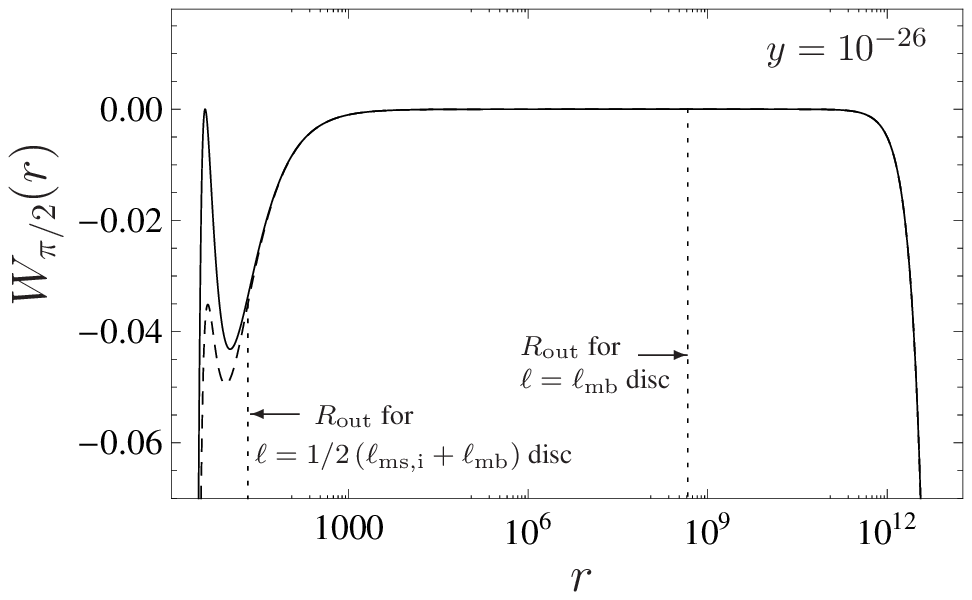}
\caption{Behaviour of the relativistic potential $W$ in the equatorial plane of the SdS spacetime with $y=10^{-26}$ for two values of the specific angular momentum: $\ell=\ell_{\rm mb}\doteq 4\,M$ (solid) and $\ell=1/2\,(\ell_{\rm ms,i}+\ell_{\rm mb})\doteq 3.8\,M$ (dashed). Notice the extremely flat plato extended around static radius $r_{\rm s}\sim y^{-1/3}$. Extension of such a plato grows with $y$ descending.}
\label{f12}
\end{figure}

For $\ell_{\rm ms,i}<\ell=\mbox{const}<\ell_{\rm mb}$, both the extension of the accretion disc (toroidal structure with the inner cusp) and the central potential depths $\Delta W_{\rm c}$ (or $\Delta w_{\rm c}$) are smaller as compared with the most extended marginally bound accretion discs. Thus the central temperature is also smaller. On the other hand, when the condition $m=M/10$ is assumed, the mass-density and pressure of the accretion tori become higher as compared to the most extended discs. The extension of the toroidal structure falls sharply with a small descend   of the $\ell=\mbox{const}$ value from $\ell_{\rm mb}$. On the other hand, it falls slowly while $\ell=\mbox{const}$ approaches $\ell_{\rm ms,i}$, as shown in the case of the most massive black hole in the table \ref{t4}. This is caused by the behaviour of the potential $W_{\pi/2}(r)$ which contains an extremely extended and flat plato near the static radius $r_{\rm s}$ and deep valley near the inner cusp (figure \ref{f12}). When the specific angular momentum $\ell=\mbox{const}\rightarrow \ell_{\rm ms,i}$, the extension of accretion torus shrinks to zero being concentrated at $r_{\rm ms,i}$. 

It is useful to show behaviour of central values of the mass-density and pressure in the case of accretion tori restricted to vicinity of the central black hole, i.e., with $\ell=\mbox{const}$ close to $\ell_{\rm ms,i}$. In fact, such tori are obtained with a moderate case of the representative value of $\ell \in (\ell_{\rm ms,i},\ell_{\rm mb})$, namely $\ell=1/2(\ell_{\rm ms,i}+\ell_{\rm mb})$, see table~\ref{t2}. Here, we have included also the astrophysically very important case of stellar mass black holes, having a typical mass  $M=10\,{\rm M}_{\odot}$, as frequently observed in Galactic X-ray binaries \cite{Rem-McCli:2006:ARASTRA:}.

Notice that for such values of $y$ and $\ell$, the extension of the tori in units of BH mass $M$ is almost independent of the cosmological parameter $y$. However, the central mass-density and pressure depend strongly on $y$ and only slightly on the adiabatic index $n$. Contrary to the case of most extended tori, the accretion tori located close to the BH horizon reflect strong relativistic phenomena. Therefore the predictions of the PN and GR models differ for more than $10\%$ and up to $40\%$. The chosen value of $\ell=1/2(\ell_{\rm ms,i}+\ell_{\rm mb})=\mbox{const}$ corresponds to tori located close to the BH horizon in regions where the influence of the cosmological constant is very small; extension of these tori is again small ($R_{\rm out}\doteq 20.6\,M$). Since we use the limiting mass of the disc $m=M/10$, central mass-density and pressure reach high values, which  grow enormously with lowering mass of the central black hole. For $M=10 {\rm M}_{\odot}$, i.e., a typical  stellar mass black hole, the central density of the torus reaches the values for which the neutrons start to drip out of the nuclei \cite{Bay-Pet-Sut:1971:ASTRJ2:}.

\subsubsection{Excretion tori}
\begin{figure}
\centering
\includegraphics[width=0.75\hsize]{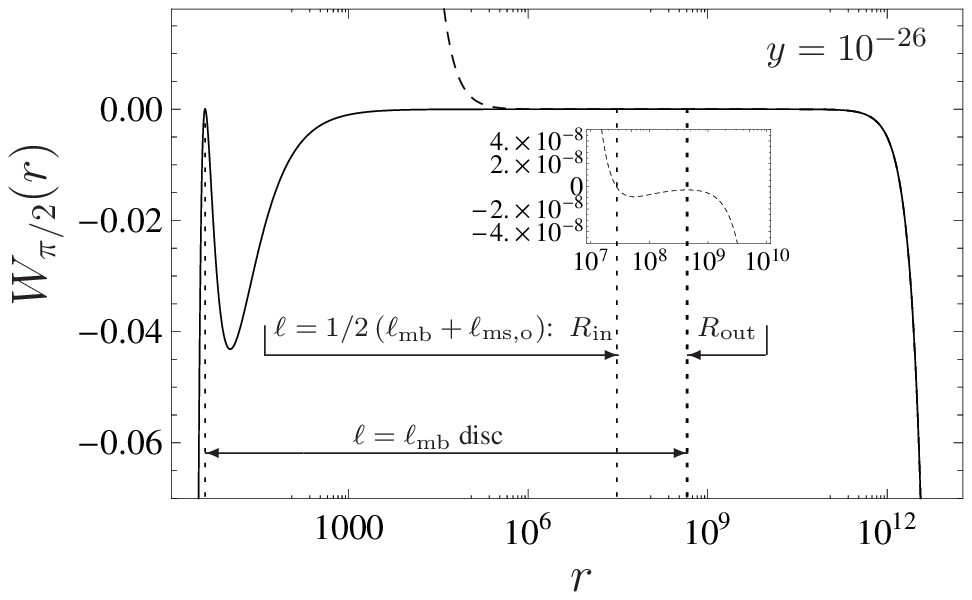}
\caption{Behaviour of the relativistic potential $W$ in the equatorial 
plane of the SdS spacetime with $y=10^{-26}$ for two values of the 
specific angular momentum: $\ell=\ell_{\rm mb}\doteq 4.0M$ (solid) and 
$\ell=1/2\,(\ell_{\rm mb}+\ell_{\rm ms,o})\doteq 7.4\times 10^3M$ (dashed), representing the most extended torus and a typical low density excretion torus.}
\label{f13}
\end{figure}

For $\ell_{\rm mb}<\ell=\mbox{const}<\ell_{\rm ms,o}$, both the extension of the excretion tori (toroidal structures with the outer cusp) and their central potential depths are again lowered in comparison to the most extended tori for fixed BH mass $M$, and the central temperature does not reach values of marginally bound discs. Contrary to the case of accretion tori, the central values of mass-density and pressure of the excretion tori are also lowered relative to the most extended tori. Extension of excretion tori falls very sharply when $\ell=\mbox{const}$ slightly overcomes $\ell_{\rm mb}$ (due to radical shift of the inner edge), while it slowly shrinks to zero extension at $r_{\rm ms,o}$ when $\ell=\mbox{const}\rightarrow \ell_{\rm ms,o}$ (see table \ref{t5} constructed for the most massive black hole with $y=10^{-26}$). Again, this is caused by the very flat and extended plato of the potential $W_{\pi/2}(r)$ near the static radius (see figure \ref{f13}). While $\ell=\mbox{const}$ grows for excretion tori of fixed mass $m=M/10$, radius of their centre in the symmetry plane grows, while their section perpendicular to the symmetry plane shrinks. The central density falls with $\ell=\mbox{const}$ growing for large part of the range of $\ell \in (\ell_{\rm mb}, \ell_{\rm ms,o})$, but it starts to grow again as $\ell=\mbox{const}$
is close enough to $\ell_{\rm ms,o}$, reaching high values especially for $\ell \rightarrow \ell_{\rm ms,o}$, since the perpendicular section of the tori shrinks to zero in such a situation (see table \ref{t5}).

We demonstrate behaviour of the mass-density and pressure for excretion tori with a moderate value of $\ell=\mbox{const}=1/2(\ell_{\rm mb}+\ell_{\rm ms,o})$. Such a value of $\ell=\mbox{const}$ corresponds to a torus with rather extreme properties; notice the values of the critical angular momentum for $y=10^{-26}$, i.e., $\ell_{\rm ms,i}\doteq 3.7\,M$, $\ell_{\rm mb}\doteq 4\,M$ and $\ell_{\rm ms,o}=1.5\times 10^4\,M$. The results are given in table \ref{t2}. The excretion tori are discussed in the astrophysically relevant cases of supermassive black holes, when extension of such a toroidal structure is comparable to the related galaxy extension.\footnote{Recall that for intermediate or stellar-mass black holes in star clusters or binary systems, the mentioned toroidal configurations are much more extended in comparison with the clusters or binary systems.}

The central mass-density in such excretion tori is in many orders lower than those of the most extended tori (or the accretion tori) having the same mass $m=M/10$. It is almost independent of the cosmological parameter $y$, and thus on the BH mass, contrary to the other two cases, where the differences are in many orders. 
The excretion tori with such a moderate value of $\ell=\mbox{const}$ are located in regions close to the static region, where the PN approach can be applied with an extremely high precision - in fact, the PN and GR predictions differ at the $7$-th (or higher) decimal order. Further, the central densities and pressure are almost independent of the adiabatic index $n$ for such distant excretion tori. 

\begin{table}
\caption{Position of the inner edge $R_{\rm in}$ of excretion tori in units of $M$ in dependence on the value of constant specific angular momentum (the results are the same for both PN and GR marginally stable tori) and the GR values (of the same order as PN values) of central densities in $\rm kg/m^3$ for $n=1.5$ in the disc which is characterized by parameter $k$ in the relation $\ell=\ell_{\rm ms,o}-k(\ell_{\rm ms,o}-\ell_{\rm mb})$ and value of cosmological parameter $y=10^{-26}$. Note that there is no significant dependence of the outer edge position on the parameter $k$, being located at $R_{\rm out}\sim 10^8\,M$.}
\begin{indented}
\item[]\begin{tabular}{@{}llllllll}
\br
$k$ &$0.1$&$0.3$&$0.5$&$0.9$&$0.99$&$0.9999$&$1$\\
\mr
$R_{\rm in} {\rm [M]}$&$10^8$&$10^8$&$10^7$&$10^6$&$10^4$&$10^1$&$10^0$\\
$\rho_{\rm c} {\rm [kg/m^3]}$&$10^{-25}$&$10^{-26}$&$10^{-25}$&$10^{-24}$&$10^{-21}$&$10^{-16}$&$10^{-16}$\\
\br
\end{tabular}
\end{indented}
\label{t5}
\end{table}

\subsection{On definition of pseudo-Newtonian potential}
Some astrophysical applications of exact GR methods represent samples of too complicated calculations. Therefore, approximative PN or even Newtonian methods are usually used in order to obtain more or less satisfactory results.    
There is a variety of PN approaches based on different definitions of the gravitational potential, describing some aspects of BH physics \cite{Pac-Wii:1980:ASTRA:,Muk:2002:ASTRJ2:,Now-Wag:1991:ASTRJ2:,Cha-Kha:1992:MONNR:,Art-Bjo-Nov:1996:ASTRJ2:,Sem-Kar:1999:ASTRA:}.  

In study of accretion discs, the most convenient and widely used is the Paczy\'{n}ski-Wiita  gravitational potential \cite{Pac-Wii:1980:ASTRA:}. Originally, this potential was introduced by a~guess, when attempting to include the Schwarzschild radius into the Newtonian gravity. However, there is a simple heuristic method for derivation of the PN potential based on the properties of the test-particle motion, giving in the case of Schwarzschild spacetime just the Paczy\'{n}ski-Wiita potential and used in derivation of the PN gravitational potential for the equatorial plane of the Kerr spacetime as well \cite{Muk:2002:ASTRJ2:}. In this paper, we used this method to define the PN potential $\psi$ for the SdS spacetimes. 

Looking for a definition of PN potential from the point of view of fluid tori description in spherically symmetric spacetimes, one can state the following. The Newtonian and GR versions of potentials, determining equipressure surfaces in barotropic marginally stable tori, are given by 
\begin{eqnarray}
w_{\rm N}=\psi_{\rm N}+\frac{l^2}{2r^2\sin^2\theta},                            \label{e71} \\
W=\ln\left(\frac{-g_{tt}g_{\phi\phi}}{\ell^2 g_{tt}+g_{\phi\phi}}\right)^{1/2}, \label{e72}
\end{eqnarray} 
where $\psi_{\rm N}$ and $l$ are the Newtonian gravitational potential and angular momentum per unit mass, while $\ell=L/E$ is the relativistic specific angular momentum. For marginally stable tori, both $l$ and $\ell$ are constant. Intending to modify the potential $w_{\rm N}$ in such a~way that its modified version $w$ corresponds to the same family of equipotential surfaces as the general relativistic potential $W$, we have to require
\begin{equation}                                                                \label{e73}
\frac{\partial_r W}{\partial_{\theta}W}=\frac{\partial_r w}{\partial_{\theta}w}.
\end{equation}
A natural modification of relation (\ref{e71}) is based on replacement of the Newtonian gravitational potential $\psi_{\rm N}$ by a~new PN potential $\psi$ which fulfills condition (\ref{e73}).

In the SdS spacetime, the GR potential $W$ is given by relation (\ref{e20}). Because of the spherical symmetry of the spacetime, we assume $\psi\neq\psi(\theta)$ and, applying the condition (\ref{e73}), we obtain an expression for the radial gradient of the PN potential $\psi$ in the form:
\begin{equation}                                                                \label{e74}
\p_r\psi=\frac{l^2(1-yr^3)}{\ell^2(r-2-yr^3)^2}.
\end{equation}
Now, it is clear that it is necessary to identify the Newtonian angular momentum per unit mass $l$ with the GR specific angular momentum $\ell=L/E$ to guarantee that the potential $\psi$ would be independent of the constant of motion, just as considered when defining $\psi$ by relation (\ref{e1}) in the general heuristic method. Then, of course, the integration of equation (\ref{e74}) yields $\psi$ in the form (\ref{e9a}), i.e., identical with the heuristic method inspired by the test-particle motion.  

\section{Conclusions}
\label{con}
It was shown that the PN gravitational potential $\psi$ defined for SdS spacetimes in close analogy with the well known Pacz\'{n}ski-Wiita potential \cite{Pac-Wii:1980:ASTRA:} reflects existence of the static radius, diverges at both the black-hole and cosmological horizons, and predicts locations of both the inner and outer marginally stable and marginally bound circular orbits at the same radii as those following from the full GR theory \cite{Stu-Kov:2008:INTJMD:}. The energy difference between the inner and outer marginally stable circular orbit, which plays a crucial role in the theory of thin discs, is very close to the relativistic result (for more details and comparison to the geodesic motion see \cite{Stu-Kov:2008:INTJMD:}). Here, we have demonstrated that the PN potential $\psi$ can be well applied even in the case of thick discs orbiting SdS black holes. It provides exact determination of the equipressure (equipotential) surfaces governing the shape of  toroidal discs in equilibrium configuration; to be precise, the shape of equipotential surfaces of the $\ell = \mbox{const}$ tori are given by identical formulae when spherical coordinates are used in the PN approach and standard Schwarzschild coordinates in the GR approach. 
In a given SdS background, the marginally stable toroidal configuration with $\ell=\ell_{\rm mb}=\mbox{const}$ is the most extended one. For $y\lesssim 10^{-10}$ extension of marginally bound tori, expressed in terms of the gravitational radius $r_{\rm h}\sim M$, falls with mass parameter $M$  according to the law $R_{\rm out}=r_{\rm mb,o}\sim M^{-2/3}$. For $\ell=\mbox{const}\lesssim \ell_{\rm mb}$, extension of the toroidal accreting configurations fastly falls even with extremely small decrease of $\ell$, similarly as the extension of excretion tori for extremely small increase of $\ell$ above $\ell_{\rm mb}$.   

We have shown that in both the PN and GR approaches there are very small differences when calculating potential barriers between the centre of the torus with $\ell=\ell_{\rm mb}=\mbox{const}$ and both its cusps. This indicates that physical properties of tori, predicted in both GR and PN approaches, could be very similar.  

We have tested physical properties of $\ell=\mbox{const}$ (marginally stable) toroidal discs in the case of adiabatic tori with fluid equation of state corresponding to the ideal gas. We have shown that the profiles of thermodynamic quantities (mass-density, temperature, pressure) are given by the profiles of  equipotential surfaces, and are almost identical in both the PN and GR approaches. Some differences appear in the central part of the tori, where temperature, mass-density and pressure given by the PN model are from a few percent to tens percent higher as compared with the GR model for current  astrophysically relevant SdS spacetimes with $y<10^{-25}$. The central temperature is given by the potential depth $\Delta W = W_{\rm in}- W_{\rm c}$ only. On the other hand, the central mass-density (and pressure) is related the total mass of the disc. In the most extended discs with $\ell=\ell_{\rm mb}$, the PN and GR total masses of the disc are nearly equal, when a fixed central density is assumed, being a few percent higher in the GR model. Further, the non-relativistic fluid approximation can be conveniently  applied for description of marginally bound adiabatic tori, since the total energy density exceeds the rest energy density in restricted central parts of the tori for a few percent only.

Note that for $y\rightarrow 0$, extension of the $\ell=\ell_{\rm mb}=\mbox{const}$ tori diverges and accretion tori with $\ell<\ell_{\rm mb}$ are relevant only. Precision of the PN model with $\ell=\mbox{const}<\ell_{\rm mb}$ grows with $\ell \rightarrow \ell_{\rm mb}$. Therefore the results of the PN approach in Schwarzschild spacetimes \cite{Muk:2002:ASTRJ2:} can be applied for adiabatic tori.

Extension of accretion (excretion) tori rapidly shrinks to the regions centered around the marginally  stable circular geodesics $r_{\rm ms,i}$ ($r_{\rm ms,o}$) when the torus parameter $\ell=\mbox{const}$ descends (rises) from the critical value $\ell=\ell_{\rm mb}$.

In the moderate cases of accretion (excretion) tori, around currently astrophysically relevant SdS black holes,  defined by the condition $\ell=\mbox{const}=1/2(\ell_{\rm ms,i}+\ell_{\rm mb})$ ($\ell=\mbox{const}=1/2(\ell_{\rm mb}+\ell_{\rm ms,o})$), with the fixed limiting mass $m=M/10$, significant differences appear. We find the central mass-density and pressure to be highly dependent on the black hole mass $M$ and adiabatic index $n$ for accretion tori; the differences of the central density and pressure obtained by the GR and PN models represent tens of percent, i.e., substantially more in comparison with the most extended tori. On the other hand, their central temperature is almost independent of the black-hole mass $M$. For excretion tori, the central mass-density is almost independent of the mass of supermassive black holes ($M \geq 10^6\,{\rm M_{\odot}}$), while their central pressure and central temperature depend significantly on $M$, because of dependence of the potential depth on $M$ in such excretion tori. The GR and PN models give quite negligible differences for these central values of thermodynamic quantities, not overcoming $10^{-4}$ percent. This behaviour can be intuitively understood easily since the accretion tori under consideration are completely located in the strong gravity region near the black-hole horizon, while the excretion tori are in the region close to the static radius where the gravity is very weak and can be very well approximated by the PN potential. The most extended $\ell=\ell_{\rm mb}$ tori have only a small part located in the strong gravity region.

We conclude that the PN modelling of tori in SdS spacetimes works very well in current astrophysically relevant situations with black holes of mass $M\lesssim 10^{10}$ M$_{\rm \odot}$ (corresponding to the cosmological parameter $y\lesssim 10^{-26}$) for tori with $\ell$ close to $\ell_{\rm mb}$ and for all excretion discs with $\ell_{\rm mb}<\ell<\ell_{\rm ms,o}$, since physical properties of the fluid are properly reflected in these situations by the PN approximation. The PN approach can be useful even for the accretion tori in strong gravity regions, but the relativistic phenomena can shift the GR results for thermodynamic quantities by tens percent of PN results. 

We believe that the PN approach is useful in describing a wide range of astrophysical phenomena influenced by the observed relic cosmological constant. Especially, we can assume that the PN potential could be useful in developing models of excretion discs that could (potentially) serve as seeds of dark matter halos. 

\section*{Acknowledgment}
This work was done as a~part of the reseach project MSM~4781305903.

\bibliographystyle{unsrt} 
\section*{References}


\end{document}